\documentclass[12pt,a4paper]{article}
\usepackage{amsmath}
\usepackage{amsfonts}
\usepackage{amssymb}
\usepackage{textcomp}
\usepackage[dvips]{graphicx}
\usepackage{epsfig}
\usepackage{bm}
\usepackage{dcolumn}
\usepackage{amsfonts}
\usepackage{color}
\usepackage[left=2cm,right=2cm,top=2cm,bottom=2cm]{geometry}
\begin{document}

\title{\textbf{Primordial black holes production in Einstein-Gauss-Bonnet gravity}}
\author{ Ekaterina~O.~Pozdeeva\footnote{E-mail address:
pozdeeva@www-hep.sinp.msu.ru} \ and \ Sergey~Yu.~Vernov\footnote{E-mail address:
svernov@theory.sinp.msu.ru} \\
\small  Skobeltsyn Institute of Nuclear Physics, Lomonosov Moscow State University,\\
\small Leninskie Gory 1, Moscow  119991,  Russia}
\date{ \ }
\date{ \ }
\maketitle
\begin{abstract}
We consider models with the Gauss-Bonnet term multiplied to a function of the scalar field and propose a new inflationary model which provides a mechanism for the formation of primordial black holes (PBHs). For certain values of the model parameters, the predictions of this model agree with observational data from the Atacama Cosmology Telescope (ACT). Due to estimation of possiable PBHs masses, we demonstrate that they can be considered as a possible source for all or part of dark matter.
\end{abstract}

\section{INTRODUCTION}

There are several reasons to include the Gauss-Bonnet term multiplied by a function of a scalar field in the action.
During cosmological inflation, the spacetime curvature is high, so it is natural to include not only the standard Einstein-Hilbert action, but also higher-order curvature terms. Such terms have been derived from low-energy effective actions in string theory~\cite{Metsaev:1986yb}. The Gauss-Bonnet scalar arises in string theory as a quantum correction to the Einstein-Hilbert action and is a natural ingredient of string-inspired cosmological models~\cite{Berkin:1990nu,Kawai:1999pw,Cartier:2001is,Hwang:2005hb,Nojiri:2005vv,Sami:2005zc,Tsujikawa:2006ph,Cognola:2006sp,Guo:2006ct,Iihoshi:2007vv,
RomeroCastellanos:2018inv,Ketov:2022lhx,Ketov:2022zhp,Gitsis:2023cyw,Odintsov:2026cxz}.

Inflationary models with the Gauss-Bonnet term are actively developed~\cite{Guo:2010jr,Jiang:2013gza,Koh:2014bka,Hikmawan:2015rze,vandeBruck:2015gjd,vandeBruck:2017voa,Koh:2016abf,Mathew:2016anx,Chakraborty:2018scm,Yi:2018gse,Odintsov:2018zhw,Kleidis:2019ywv,
Odintsov:2019clh,Odintsov:2020zkl,Odintsov:2020sqy,Rashidi:2020wwg,Pozdeeva:2020shl,Pozdeeva:2020apf,Pozdeeva:2021iwc,Oikonomou:2021kql,Pozdeeva:2021nmz,Kawai:2021edk,Odintsov:2022zrj,Oikonomou:2022ksx,Kawaguchi:2022nku,
Gangopadhyay:2022vgh,Venikoudis:2022gfg,Odintsov:2023aaw,Nojiri:2023mvi,Ashrafzadeh:2023ndt,Odintsov:2023lbb,Mudrunka:2023wxy,Koh:2023zgn,Nojiri:2024cal,Pozdeeva:2024kzb,Pozdeeva:2024ihc,Oikonomou:2024etl,Ashrafzadeh:2024oll,Solbi:2024zhl,Odintsov:2025wai,Mudrunka:2025xcg,Ahghari:2025hfy,
Yogesh:2025hll,Yogesh:2025wak,Aldabergenov:2025oys,Rudenko:2026wkb,Yogesh:2026esn,Oikonomou:2026qkj,Manucharyan:2026twj,Odintsov:2026veu}.
In many of these models, the propagation speeds of scalar and tensor perturbations are different from the speed of light only during the very early stages of the Universe evolution.
For this reason, such inflationary models are consistent with the GW170817 observations~\cite{LIGOScientific:2017ync} without the need for additional assumptions\footnote{Note that the authors of some Einstein-Gauss-Bonnet inflation models~\cite{Odintsov:2019clh,Odintsov:2020zkl,Odintsov:2020sqy,Odintsov:2022zrj,Venikoudis:2022gfg,Oikonomou:2022ksx,Odintsov:2023aaw,Oikonomou:2024etl,Odintsov:2025wai,Oikonomou:2026qkj,Odintsov:2026veu}  require the speed of gravitational waves to be equal to the speed of light during inflation.}. On the other hand, instabilities associated with negative squared propagation speeds may arise, so, one needs to check the stability conditions are satisfied during inflation. Models with the Gauss-Bonnet term multiplied by functions of a scalar field form a subset of Horndeski theories~\cite{Kawaguchi:2022nku,Kobayashi:2011nu}. So, these models are governed by second-order evolution equations. It ensures the absence of the Ostrogradsky instability.

The more recent observational data from ACT/DESI collaborations~\cite{AtacamaCosmologyTelescope:2025blo,AtacamaCosmologyTelescope:2025nti,DESI:2024mwx,DESI:2025zgx} slightly shift the tolerated range of values for $n_s$:
\begin{equation}\label{ACT}
n_s = 0.9743 \pm 0.0034\,.
\end{equation}

The ACT/DESI data are consistent with the previously obtained upper bounds on the  tensor-to-scalar perturbation ratio $r$ and value of the scalar power spectrum amplitude $A_s$~\cite{Planck:2018jri,BICEP:2021xfz,Galloni:2022mok}:
\begin{equation}
\label{Asrobserv}
A_s=(2.10\pm 0.03)\times 10^{-9}\qquad {\rm and} \qquad  r < 0.028\,.
\end{equation}

Primordial black holes (PBHs) can play an important role in cosmology. They can be considered as possible candidates for dark matter~\cite{Ivanov:1994pa,Khlopov:2008qy,Kamenshchik:2018sig,Carr:2020xqk,Green:2020jor,Carr:2020gox,Ozsoy:2023ryl}.
The current observations support the hypothesis that PBHs are possible seeds the supermassive black holes, as it is difficult to explain the existence of some of these objects through models of stellar collapse~\cite{Dolgov:2017aec,Dolgov:2020sov}. Also, PBHs can be responsible for some observed gravitational wave events~\cite{Sasaki:2018dmp,DeLuca:2025fln}.

The presence of overdensities exceeding a critical value during the cosmological inflation is necessary for the production of PBHs during the radiation-dominated epoch.
Inflation can create these overdensities, but slow-roll inflationary models cannot generate them~\cite{Dolgov:1992pu,Ivanov:1994pa,Garcia-Bellido:1996mdl,Ketov:2021fww,Ozsoy:2023ryl}. In single-field models, including those with the Gauss-Bonnet term, the formation of PBHs is connected with the ultra-slow-roll stage during inflation~\cite{Garcia-Bellido:2017mdw,Germani:2017bcs,Kannike:2017bxn,Ezquiaga:2017fvi,Kristiano:2022maq,Chataignier:2023ago,Kamenshchik:2024kay,Saburov:2023buy,Saburov:2024und,Ketov:2025cqg,Iovino:2025tcv,Totolou:2025dsb}.

The PBH production in inflation models with the Gauss-Bonnet term has been investigated in Refs.~\cite{Kawai:2021edk,Kawaguchi:2022nku,Ashrafzadeh:2023ndt,Ashrafzadeh:2024oll,Solbi:2024zhl,Yogesh:2025hll}.
These models are in good agreement with the Planck/BICEP observations of the
cosmic microwave background (CMB) radiation~\cite{Planck:2018jri,BICEP:2021xfz}, but have tensions with the ACT data~\cite{AtacamaCosmologyTelescope:2025blo,AtacamaCosmologyTelescope:2025nti}.
So, it is actual to construct a new inflation models with the Gauss-Bonnet term that is in agreement with the ACT data and is suitable for PBH production.
As known the existence of the ultra-slow-roll stage during inflation makes the evolution of the scalar field slowly that
 decreases the value of $n_s$ calculated at a fixed value of e-folding before the end of inflation~\cite{Kawaguchi:2022nku}.
 Thus, models with primordial black hole (PBH) production are in stronger contradiction with the ACT data than inflationary models that do not include PBHs. By this reason, we do not use the standard way of construction of inflationary models with PBH production by adding the Gauss-Bonnet term to the known model without Gauss-Bonnet term that do not include PBHs and is in good agreement with the current observation data.

In this paper, we construct a model with such potential that the inflationary stage is too short if the Gauss-Bonnet term is not present. We show that the adding the Gauss-Bonnet term multiplied by the suitable function of the scalar field gives the inflation model with PBH production that agrees with the ACT observational data. We estimate the PBH masses in the model constructed to demonstrate that PBHs could be considered dark matter candidates.

The paper is organized as follows. In section 2, we remind the evolution equations of the considered model. In section 3, we describe the slow-roll and ultra-slow-roll approximation of the evolution equations in terms the effective potential.
In section 4, we construct a model with the Gauss-Bonnet term that satisfies the ACT observation data and can describe the PBH production. We summaries our results in section 5.

\section{EINSTEIN-GAUSS-BONNET GRAVITY MODELS}

We consider the  model, described by the following action:
\begin{equation}
\label{action1}
S=\int d^4x\sqrt{-g}\left[\frac{M_\mathrm{Pl}^2}{2}R-\frac{g^{\mu\nu}}{2}\partial_\mu\phi\partial_\nu\phi-V(\phi)-\frac{\xi(\phi)}{2}{\cal G}\right],
\end{equation}
where  $R$ is the Ricci scalar,  $M_\mathrm{Pl}$ is the reduced Planck mass, $V(\phi)$ and $\xi(\phi)$ are differentiable functions, and the Gauss-Bonnet term
\begin{equation}\label{GBterm}
{\cal G}=R_{\mu\nu\rho\sigma}R^{\mu\nu\rho\sigma}-4R_{\mu\nu}R^{\mu\nu}+R^2.
\end{equation}

In the spatially flat Friedmann-Lema\^{i}tre-Robertson-Walker metric with
\begin{equation*}
ds^2={}-dt^2+a^2(t)(dx^2+dy^2+dz^2),
\end{equation*}
one obtains the following system of evolution equations~\cite{vandeBruck:2015gjd,Pozdeeva:2020apf}:
\begin{eqnarray}
6M_\mathrm{Pl}^2H^2&=&\psi^2+2V+24\xi_{,\phi}\psi H^3,\label{Equ00} \\
2M_\mathrm{Pl}^2\dot{H}&=&{}-\psi^2+4H^2\left(\xi_{,\phi\phi}\psi^2+\xi_{,\phi}\dot{\psi}\right)+4H\xi_{,\phi}\psi \left(2\dot{H}-H^2\right),
\label{Equ11} \\
\dot\phi&=&\psi,\\
\dot{\psi}&=&{}-3H\psi -V_{,\phi} -12H^2\xi_{,\phi}\left(\dot{H}+H^2\right),\label{Equphi}
\end{eqnarray}
where $H=\dot{a}/a$ is the Hubble parameter, $a(t)$ is the scale factor, dots denote the derivatives with respect to the cosmic time $t$ and $A_{,\phi} \equiv \frac{dA}{d\phi}$ for any function $A(\phi)$.

Equations~(\ref{Equ11})---(\ref{Equphi}) are equivalent to the following dynamical system~\cite{Pozdeeva:2019agu,Pozdeeva:2024ihc}:
\begin{equation}
\label{DynSYS}
\begin{split}
\dot\phi=&\psi,\\
\dot\psi=&\frac{1}{B-2\xi_{,\phi}H\psi}\left\{
3\left[3-4\,\xi_{,\phi\phi} H^2 \right]\xi_{,\phi}H^2\psi^2+\left[3B+2\,\xi_{,\phi}V_{,\phi}-3M_\mathrm{Pl}^2\right]H\psi-\frac{2V^2}{M_\mathrm{Pl}^2}X\right\},\\
\dot H=&\frac{1}{4\left(B-2\,\xi_{,\phi}H\psi\right)}\left\{\left(4\,\xi_{,\phi\phi}H^2-1\right)\psi^2-16\xi_{,\phi}H^3\psi-16\frac{V^2}{M_\mathrm{Pl}^4}\,\xi_{,\phi} H^2 X\right\},
\end{split}
\end{equation}
where
\begin{equation*}
B=12\xi_{,\phi}^2H^4+\frac{M_\mathrm{Pl}^2}{2},\qquad
X=\frac{M_\mathrm{Pl}^4}{4V^2}\left(12\xi_{,\phi} H^4+V_{,\phi}\right).
\end{equation*}

Using the e-folding number $N=\ln(a/a_{b})$, where $a_{b}$ is a
constant, as an independent variable and introducing $\chi=\psi/H$, we rewrite
system~(\ref{DynSYS}) as follows:
\begin{equation}
\label{DynSYSN}
\begin{split}
\phi'=&\,\chi,\\
\chi'=&\,\frac{1}{H^2\left(B-2\xi_{,\phi}H^2\chi\right)}\left\{
3\left[3-4\xi_{,\phi\phi} H^2\right]\xi_{,\phi}H^4\chi^2+ \left[3B+2\xi_{,\phi}V_{,\phi}-3M_\mathrm{Pl}^2\right]H^2\chi-\frac{2V^2}{M_\mathrm{Pl}^2}X\right\}\\
&{}-\frac{\chi}{4\left(B-2\xi_{,\phi}H^2\chi\right)}\left\{\left(4\xi_{,\phi\phi}H^2-1\right)\chi^2-16\xi_{,\phi}H^2\chi-16\frac{V^2}{M_\mathrm{Pl}^4}\xi_{,\phi}X\right\},\\
\left(H^2\right)'=&\,\frac{H^2}{2\left(B-2\xi_{,\phi}H^2\chi\right)}\left\{\left(4\xi_{,\phi\phi}H^2-1\right)\chi^2-16\xi_{,\phi}H^2\chi-16\frac{V^2}{M_\mathrm{Pl}^4}\xi_{,\phi}X\right\},
\end{split}
\end{equation}
where primes denote the derivatives with respect to $N$.

Equation (\ref{Equ00}) can be presented in the following form
\begin{equation}
\label{Equ00N}
24\chi\xi_{,\phi} H^4+\left(\chi^2- 6M_\mathrm{Pl}^2\right)H^2+2V=0.
\end{equation}

Equation (\ref{Equ00N}) restricts the set of initial conditions of system~(\ref{DynSYSN}).

\section{SLOW-ROLL AND ULTRA-SLOW-ROLL APPROXIMATIONS}

Using Eq.~\eqref{Equ00}, we can rewrite system~(\ref{DynSYS}) into a form that is more suitable for the slow-roll and ultra-slow-roll approximations.
In these approximations, only zero-order and first-order terms in $\psi$ are used on the right-hand side of system~(\ref{DynSYS}).

Substituting
\begin{equation}
H^2=\frac{\psi^2+2V+24H^3\xi_{,\phi}\psi}{6M_\mathrm{Pl}^2},
\end{equation}
we obtain the second equation of system \eqref{DynSYS} as follows:
\begin{equation}\label{secondequationapprox}
\begin{split}
\dot{\psi}&=\frac{1}{ \left( 3\,M_\mathrm{Pl}^{6}+8V^2\xi_{,\phi}^{2}\right)^{2}}\left[{}-3H\psi\left( 9\,M_\mathrm{Pl}^{12}+48\,M_\mathrm{Pl}^{6}  V^{2} \xi_{,\phi}^{2}-96\,M_\mathrm{Pl}^{4} V^{2} \xi_{,\phi}^{3}V_{,\phi} - 64V^{4}
\xi_{,\phi}^{4}\right)\right.\\
&\left. { } - M_\mathrm{Pl}^{2}\left( 3M_\mathrm{Pl}^{4}V_{,\phi} +4 V^{2}\xi_{,\phi}\right)\left( 3M_\mathrm{Pl}^{6}+8 V^{2}\xi_{,\phi}^{2}\right)
\right]+{\cal{O}}(\psi^2)\,.
\end{split}
 \end{equation}

Using the effective potential\footnote{We use the potential $V_{eff}(\phi)$ proposed in Ref.~\cite{Pozdeeva:2019agu} (see also Ref.~\cite{Vernov:2021hxo}). Later, a different effective potential, which depends on both $\phi$ and $H$, has been proposed in Ref.~\cite{Kawaguchi:2022nku}.}
\begin{equation}\label{Veff}
V_{eff}=\frac{1}{3}\xi-\frac{M_\mathrm{Pl}^4}{4V}\,,
\end{equation}
we substitute
  \begin{equation}\label{dxiVeff}
\xi_{,\phi}=3{V_{eff}}_{,\phi}-\frac{3M_\mathrm{Pl}^4V_{,\phi}}{4V^2}\,
\end{equation}
and get Eq.~(\ref{secondequationapprox}) in the following form:
 \begin{equation}\label{equpsiVeff}
 \begin{split}
\dot{\psi}&={}-3H\psi+\frac{216HV^{2}{V_{eff}}_{,\phi}\left(4V^2{V_{eff}}_{,\phi}-M_\mathrm{Pl}^4V_{,\phi}\right)^3 }{ \left( 3M_\mathrm{Pl}^{8} V_{,\phi}^{2}
-24M_\mathrm{Pl}^{4}{V_{eff}}_{,\phi} V^2V_{,\phi} +2M_\mathrm{Pl}^{6} V^{2}+48{V^{2}_{eff}}_{,\phi} V^{4}\right)^{2}}\,\psi\\
&-\frac{8M_\mathrm{Pl}^{2}V^{4}{V_{eff}}_{,\phi}}{3 M_\mathrm{Pl}^{8} V_{,\phi}^{2}-24 M_\mathrm{Pl}^{4} V^{2} V_{,\phi}{V_{eff}}_{,\phi} +2\,M_\mathrm{Pl}^{6} V^{2}
+48{V^{2}_{eff}}_{,\phi} V^4}+{\cal{O}}(\psi^2)\,.
\end{split}
\end{equation}

The first derivative of the effective potential ${V_{eff}}_{,\phi}=0$ at any de Sitter point~\cite{Pozdeeva:2019agu}.
In the USR regime, the expansion is very close to de Sitter one, so we neglect not only ${\cal{O}}(\psi^2)$ terms, but also all terms proportional to ${V_{eff}}_{,\phi}$
and get
\begin{equation}
\label{USRequ}
\dot{\psi}\approx {}-3H\psi\,.
\end{equation}
This equation is the same as in GR models with a minimally coupled scalar field, but the Gauss-Bonnet term changes the condition for the existence of the ultra-slow-roll regime,
because we neglect terms proportional to ${V_{eff}}_{,\phi}$ but not $V_{,\phi}$.

To consider Eqs.~(\ref{Equ00})--(\ref{Equphi}) in the slow-roll regime we use the following slow-roll parameters~\cite{Guo:2010jr,vandeBruck:2015gjd,Pozdeeva:2020apf,Odintsov:2023lbb}:
\begin{equation}
\label{epsilon}
\varepsilon_1 ={}-\frac{\dot{H}}{H^2}={}-\frac{1}{2}\frac{d\ln(H^2)}{dN},\qquad \varepsilon_{i+1}= \frac{d\ln|\varepsilon_i|}{dN},\quad i\geqslant 1,
\end{equation}
\begin{equation}
\label{delta}
\delta_1= \frac{4}{M_\mathrm{Pl}^2}\xi_{,\phi}H\psi=\frac{4}{M_\mathrm{Pl}^{2}}\xi_{,\phi}H^2\chi,\qquad \delta_{i+1}=\frac{d\ln|\delta_i|}{dN}, \quad i\geqslant 1.
\end{equation}
The slow-roll approximation requires $|\varepsilon_i|\ll1$ and  $|\delta_i|\ll 1$.

The propagation speeds in Einstein-Gauss-Bonnet gravity  have been considered in Refs.~\cite{Guo:2010jr,vandeBruck:2015gjd,Kawaguchi:2022nku}.
We use the following expressions of the scalar perturbation propagation speed
\begin{equation}
\label{cR}
c_{s}^2=1+\frac{8H^2{\dot\xi}^2\left[2\dot H\left(U_0-2H{\dot\xi}\right)+H^2\left(\ddot{\xi}- H{\dot\xi}\right)\right]}{\left(U_0-2H{\dot\xi}\right)
\left[{\dot\phi}^2\left(U_0-2H{\dot\xi}\right)+12 H^4{\dot\xi}^2\right]}
\end{equation}
and of the gravitational wave propagation speed
\begin{equation}
\label{cT}
    c_{t}^2 = 1-\frac{4\left(\ddot{\xi}-H{\dot\xi}\right)}{M_\mathrm{Pl}^2-4H{\dot\xi}}\,.
\end{equation}

In~Ref.~\cite{Guo:2010jr}, the spectral index of scalar perturbations and the tensor-to-scalar ratio are presented
in terms of slow-roll parameters:
\begin{eqnarray}
  n_s&=&1-2\varepsilon_1-\frac{2\varepsilon_1\varepsilon_2-\delta_1\delta_2}{2\varepsilon_1-\delta_1},\label{n_s_i} \\
  r&=&8|2\varepsilon_1-\delta_1|\label{r_i}.
\end{eqnarray}

 Using
\begin{equation}
\chi^2=\left(\phi^\prime\right)^2=2\varepsilon_1-\delta_1+\delta_1(\delta_2-\varepsilon_1)\label{DphiPrimeformF2},
\end{equation}
we get in the slow-roll approximation
\begin{equation}
(\phi^\prime)^2\approx2\varepsilon_1-\delta_1\,.
\label{DphiApprox}
\end{equation}

We simplify the expression for  spectral index of scalar perturbations  \eqref{n_s_i} using $\varepsilon_2=\varepsilon_1^\prime/\varepsilon_1$, $\delta_2=\delta_1^\prime/\delta_1$  and tensor-to-scalar ratio \eqref{r_i}. So, the  inflationary parameters can be presented in the  following form:
\begin{eqnarray}
  &&n_s=1-2\varepsilon_1+\frac{r^\prime}{r}  \label{n_r(phi)}\,,\\
  &&r=8|2\varepsilon_1-\delta_1|\approx8(\phi^\prime)^2\label{r(phi)}\,.\label{r(epsilon,delta)}
\end{eqnarray}
We get opportunity for PBH creation after inflation due to violation
of the slow-roll regime. And in our consideration it is better to
use the following formula to compute the amplitude of scalar
perturbations during inflation\cite{Granda:2019wpe}
\begin{equation}
\label{Pzeta}
{\cal
P}_{\zeta}\approx 2^{2\mu_s-3}\left(\frac{\Gamma(\mu_s)}{\Gamma(3/2)}\right)^2\frac{H^2}{8\pi^2|Q_sc_s^3|}%\mathrm{P}_s
\end{equation}
where
\begin{eqnarray}
   \mu_s &=&\frac{1}{2}+\frac{1}{1-\varepsilon_1}\left(1+\frac{1}{2}\frac{d\ln(Q_s)}{dN}\right)\,, \\
  Q_s &=& \frac{4\chi^2Q_t\left(2Q_t+12H^4\xi_{,\phi}^2\right)}{\left(M_\mathrm{Pl}^{2}-6H^2\xi_{,\phi}\chi\right)^2},\\
  Q_t &=& \frac14\left(M_\mathrm{Pl}^{2}-4H^2\xi_{,\phi}\chi\right)\,.
\end{eqnarray}

The standard expression for the scalar perturbation amplitude in the slow-roll regime~\cite{vandeBruck:2015gjd}  is
\begin{equation}
\label{As_slr} A_s\approx\frac{2H^2}{\pi^2 M_\mathrm{Pl}^2\, r}\,,
\end{equation}
which coincides with \eqref{Pzeta} at $\mu_s=3/2$. At the beginning of inflation,
all slow-roll parameters are small, so  $\mu_s\approx3/2$ and  ${\cal P}_{\zeta} \approx A_s$.

There are a few ways to get the slow-roll approximate equations.  The first variant of approximate equations has been proposed in Ref.~\cite{Guo:2010jr} and has been presented in terms of the effective potential in Ref.~\cite{Pozdeeva:2020apf}.
This way assumes that all inflationary parameters are negligibly small and can be removed directly from Eqs.~(\ref{Equ00})--(\ref{Equphi}). So, we get:
\begin{eqnarray}
H^2&\simeq&\frac{V}{3M_\mathrm{Pl}^2}\,,
\label{Equ0lo}\\
\dot H &\simeq &{}-\frac{\psi^2}{2M_\mathrm{Pl}^2}-\frac{2}{M_\mathrm{Pl}^2}\,\xi_{,\phi}H^3\psi\,, \label{EquHlo}\\
\psi& \simeq &{}-\frac{V_{,\phi}+12\xi_{,\phi}H^4}{3H}.
\label{Equphilo}
\end{eqnarray}

A new way to construct slow-roll approximate equations has been proposed in Ref.~\cite{Pozdeeva:2024ihc}. The key point is to use a dynamical system (\ref{DynSYS}) for constructing the slow-roll approximation.
These slow-roll equations are used and analyzed in~Ref.~\cite{Yogesh:2024mpa,Rudenko:2026wkb}.`
In this paper, we construct the slow-roll approximation based on Eq.~(\ref{secondequationapprox}).
In the slow-roll regime, we neglect $\dot{\psi}$ and ${\cal{O}}(\psi^2)$ terms in Eqs.~(\ref{secondequationapprox}) and (\ref{equpsiVeff}). Therefore, we get:
\begin{equation}\label{equpsislr}
\begin{split}
    H\psi&={}-\frac {M_\mathrm{Pl}^{2} \left( 3 M_\mathrm{Pl}^{4}V_{,\phi}+4V^{2}\xi_{,\phi}\right)
    \left(3M_\mathrm{Pl}^{6}+8V^{2}\xi_{,\phi}^{2}\right)}{3 \left( 9M_\mathrm{Pl}^{12}+48M_\mathrm{Pl}^{6}V^{2}\xi_{,\phi}^{2}-96M_\mathrm{Pl}^{4} V^{2}
 \xi_{,\phi}^{3}V_{,\phi}-64V^{4} \xi_{,\phi}^{4} \right) }\\
&={} -\frac{8}{3}{M_\mathrm{Pl}^{2} V^{4}V_{eff,_{\phi}} \left(3M_\mathrm{Pl}^{8} V_{,\phi}^2+2M_\mathrm{Pl}^{6} V^{2}-24M_\mathrm{Pl}^{4} V ^{2}V_{,\phi}V_{eff,_{\phi}} +48V^{4}
 V_{eff,_{\phi}}^{2} \right) }\\
&\times \left[9M_\mathrm{Pl}^{16} V_{,\phi}^{4}+12M_\mathrm{Pl}^{14} V^{2} V_{,\phi}^{2}-72M_\mathrm{Pl}^{12} V ^{2}  V_{,\phi}^{3}V_{eff,_{\phi}} +4M_\mathrm{Pl}^{12} V^{4}-96M_\mathrm{Pl}^{10} V^{4}
V_{,\phi} V_{eff,_{\phi}} \right.\\
&\left.{}+192M_\mathrm{Pl}^{6} V^{6} V_{eff,_{\phi}}^{2}+1152M_\mathrm{Pl}^{4} V^{6} V_{,\phi}V_{eff,_{\phi}}^{3}-2304V^{8}V_{eff,_{\phi}}^{4}\right]^{-1}
\equiv A(\phi)\,.
\end{split}
\end{equation}

It is important that the function $A$ depends only on $\phi$. Substituting \eqref{equpsislr} into (\ref{Equ00}) and neglecting $\psi^2$ term, we get the Hubble parameter as the following function of $\phi$:
\begin{equation}
\label{H2phi}
        H^2\approx\frac{V}{3\left(M_\mathrm{Pl}^{2}-4\xi_{,\phi}A\right)}.
\end{equation}

The slow-roll evolution of the scalar field is defined by the following first-order differential equation:
\begin{equation}
\label{chiphi}
\chi\approx\frac{A}{H^2}\approx\frac{3A}{V}\left(M_\mathrm{Pl}^{2}-4\xi_{,\phi}A\right)\approx\frac{3M_\mathrm{Pl}^{2}A}{V}\,.
\end{equation}

Using Eqs.~(\ref{H2phi}) and (\ref{chiphi}), we obtain the slow-roll parameters (\ref{epsilon}) and (\ref{delta}) as functions of~$\phi$:
\begin{equation}
\varepsilon_1\approx{}-\frac{1}{2H^2}\frac{dH^2}{d\phi}\chi\approx{}-\frac{A}{2H^4}\frac{dH^2}{d\phi},
\end{equation}
\begin{equation}
    \delta_1\approx \frac{4}{M_\mathrm{Pl}^2}\xi_{,\phi}A.
\end{equation}

\section{THE PROPOSED INFLATIONARY MODEL}

The known inflationary EGB gravity models with the PBH
production~\cite{Kawai:2021edk,Kawaguchi:2022nku,Ashrafzadeh:2024oll,Yogesh:2025hll}
have been constructed before the most recent restrictions~(\ref{ACT}) on the
inflationary parameter $n_s$ were provided by the ACT
collaboration~\cite{AtacamaCosmologyTelescope:2025blo,AtacamaCosmologyTelescope:2025nti}.

We construct a new model that is in good agreement with the ACT data. In the proposed model, the Gauss-Bonnet term is almost constant at both the beginning and the end of inflation.
It affects only a short period and violates the slow-roll regime. The appearance of the ultra-slow-roll stage of inflation provides not only the necessary condition for PBH production but also makes the duration of inflation longer.

We consider the model (\ref{action1}) with
\begin{equation}
\label{EGB model}
\begin{split}
V&=V_0\,\mathrm{e}^{\beta\left(\phi^2/M_\mathrm{Pl}^2\right)^\alpha},\\
\xi&=\frac{\xi_0}{V}\ln\left(c+(1-c)\tanh^2\left(\omega(\phi-\phi_{*})\right)\right)
\end{split}
\end{equation}
where  $V_0$, $\alpha$, $\beta$, $\omega$, $\xi_0$, $c$ and $\phi_{*}$ are numerical parameters.

At the beginning of inflation, the slow-roll approximation is carried out and we can use the appropriate formulas to
estimate parameters $V_0$, $\alpha$ and $\beta$  as well as the value of $\phi$ at which $A_s$, $n_s$, and $r$ are calculated.

Parameters $\alpha$, $\beta$ and $V_0$ allow us to get suitable values of the inflationary parameters $A_s$, $n_s$, and $r$.
Other parameters determine the ultra-slow-roll regime. To analyze full evolution during inflation we solve system (\ref{DynSYSN}) numerically without any approximation. 
It allows us to calculate the duration of inflation and possible PBH masses.

To get appropriate inflationary scenario we choose the following values for
parameters:
\begin{equation}
\label{param}
V_0=7.7354\cdot10^{-10}, \quad \beta=-1, \quad\alpha=-0.18,\quad \phi_{*}=2.4,\quad \omega=15,\quad c=3.7.
\end{equation}

For these values of parameter and $\phi=\phi_b=4.777\,M_\mathrm{Pl}$, the Hubble parameter is $H_b=1.22\cdot10^{-5}\,M_\mathrm{Pl}$ and we get the following values of the spectral index and the
 tensor-to-scalar ratio:
\begin{equation}
n_s=0.974\quad \mathrm{and}\quad r=0.0146\,.
\end{equation}

We also obtain 
\begin{equation}
A_s\approx 2.07  10^{-9} \quad \mathrm{and}\quad {\cal{P}}_{\zeta}\approx 2.10\cdot 10^{-9}.
\end{equation}

At the end of inflation, when $\varepsilon_1=1$, the Hubble parameter is $H_e=7.52\cdot10^{-6}M_\mathrm{Pl}$ and $\phi_e\approx 0.159 M_\mathrm{Pl}$.
These values are the same for all values the parameter $\xi_0$ presented in Table~\ref{Tab}. We change the parameter $\xi_0$ to get different masses of PBH (see Table~\ref{Tab}).
The point $N=0$ corresponds to $\phi=\phi_b$.  
We estimate the point $N_{*}$ as a point corresponding a jump of ${\cal{P}}_{\zeta}$ to $6$ orders.

The mass of PBHs depends on duration of the second stage of inflation, $N_e-N_{*}$, and the
value of the Hubble parameter at the end of inflation $H_{e}$.
To estimate the mass of PBHs we use the following formula~\cite{Pi:2017gih,Frolovsky:2022qpg,Pozdeeva:2024hfq}:
\begin{equation}
M_{PBH}\simeq\frac{M_\mathrm{Pl}^2}{H_{e}}\exp\left(2(N_{e}-N_{*})\right),
\end{equation}
where $N_*$ is the minimal value of $N$, at which at which the function $L\equiv\log_{10}\left({\cal P}_{\zeta}/{\cal P}_{\zeta,CMB}\right)$ is equal to $6$.

\begin{table}[h]
\begin{center}
\caption{The values of PBH mass for different values of $\xi_0$. Other model parameters are given in (\ref{param}).\label{Tab}}
\begin{tabular}{|c|c|c|c|c|c|}
  \hline
    $\xi_0$&  $N_{tot}$ & $N_{tot}-N_{*}$ & $M_{PBH}/M_\mathrm{Pl}$& $M_{PBH}/g$ & $M_{PBH}/M_{\bigodot}$  \\
   \hline
  $-0.0072165$ & $59.1$ & $19.4$ & $7.69\cdot10^{21}$ & $3.38\cdot10^{16} $ &  $1.68\cdot10^{-17}$  \\
  $-0.007217$ & $59.5$ & $19.8$ & $2.10\cdot10^{22}$ & $9.24\cdot10^{16}$& $4.58\cdot 10^{-17}$ \\
  $-0.007218$ & $60.9$ & $20.8$ & $1.04\cdot10^{23}$ & $4.58\cdot10^{17}$& $2.27\cdot10^{-16}$ \\
  $-0.0072185$ & $61.7$ & $21.7$ & $6.29\cdot10^{23}$ & $2.77\cdot10^{18}$ & $1.37\cdot10^{-15}$ \\
  $-0.007219$ & $64.0$ & $23.8$ & $5.13\cdot10^{25}$ & $2.26\cdot10^{20}$ & $1.12\cdot10^{-13}$  \\
  $-0.0072191$ & $65.1$ & $24.8$ & $3.08\cdot10^{26}$ & $1.36\cdot10^{21}$ & $6.71\cdot10^{-13}$\\
 $-0.00721915$ & $66.1$ & $25.6$ & $1.88\cdot10^{27}$ & $8.27\cdot10^{21}$& $4.10\cdot 10^{-12}$ \\
 \hline
\end{tabular}
\end{center}
\end{table}

\begin{figure}
\includegraphics[scale=0.41]{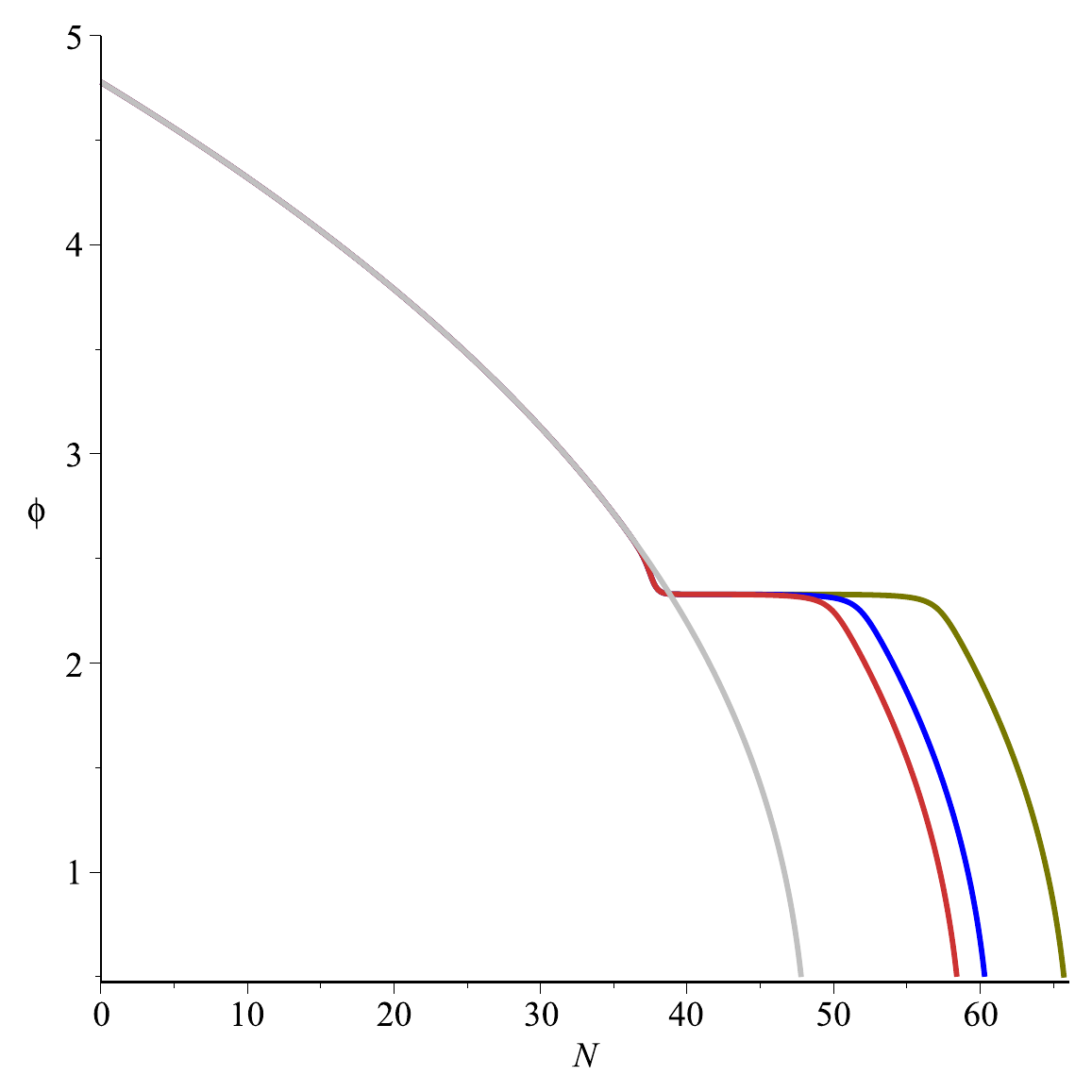}\quad
\includegraphics[scale=0.41]{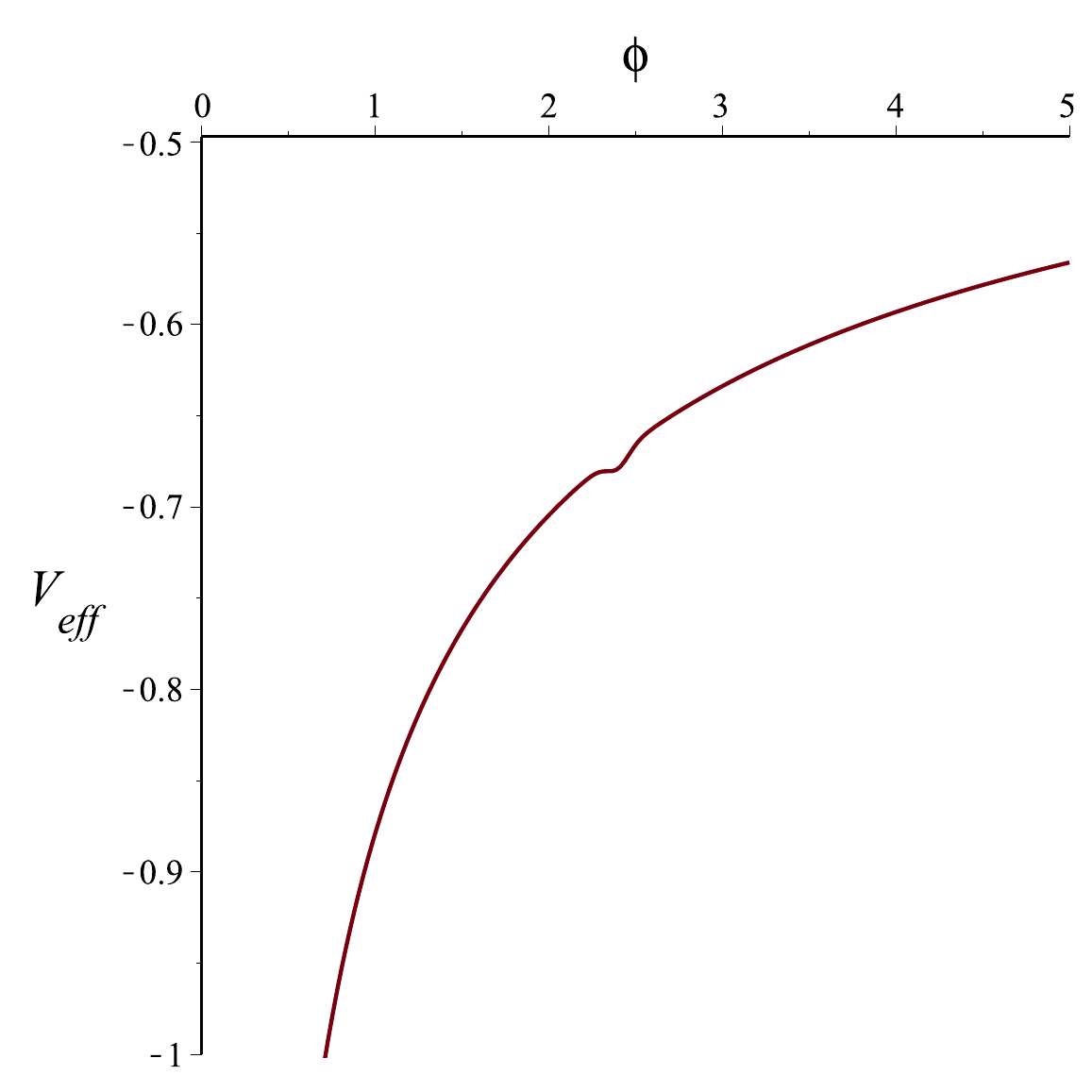}
\caption{The evolution of the scalar field  $\phi(N)$ during
inflation (left) and the effective potential $V_{eff}(\phi)$
(right) for the values of parameters given by (\ref{param}).
The evolution of  $\phi(N)$ is found by numerical integration of
system (\ref{DynSYSN}). Values of the function $\phi(N)$ is in units
of $M_\mathrm{Pl}$. The gray curve corresponds to  $\xi_0=0$,   the red curve ---
$\xi_0=-0.007216$, the
blue curve --- $\xi_0=-0.007218$, and the green one --- $\xi_0=-0.00721915$. The
effective potential $V_{eff}(\phi)$ is presented at $\xi_0=-0.00721915$. \label{Fig1}}
\end{figure}

The evolution of the scalar field $\phi$ is presented in Fig.~\ref{Fig1}. The gray curve corresponds to the model without the Gauss-Bonnet term. In this case, the inflationary stage is too short. By adding the Gauss-Bonnet term, we obtain both PBH production and a suitable duration of inflation (see the left panel of Fig.~\ref{Fig1}). One can see, that $\phi(N)$ is almost a constant near $\phi_{*}$. At this point, the effective potential changes very slowly (see the right panel of Fig.~\ref{Fig1}).

\begin{figure}
\includegraphics[scale=0.41]{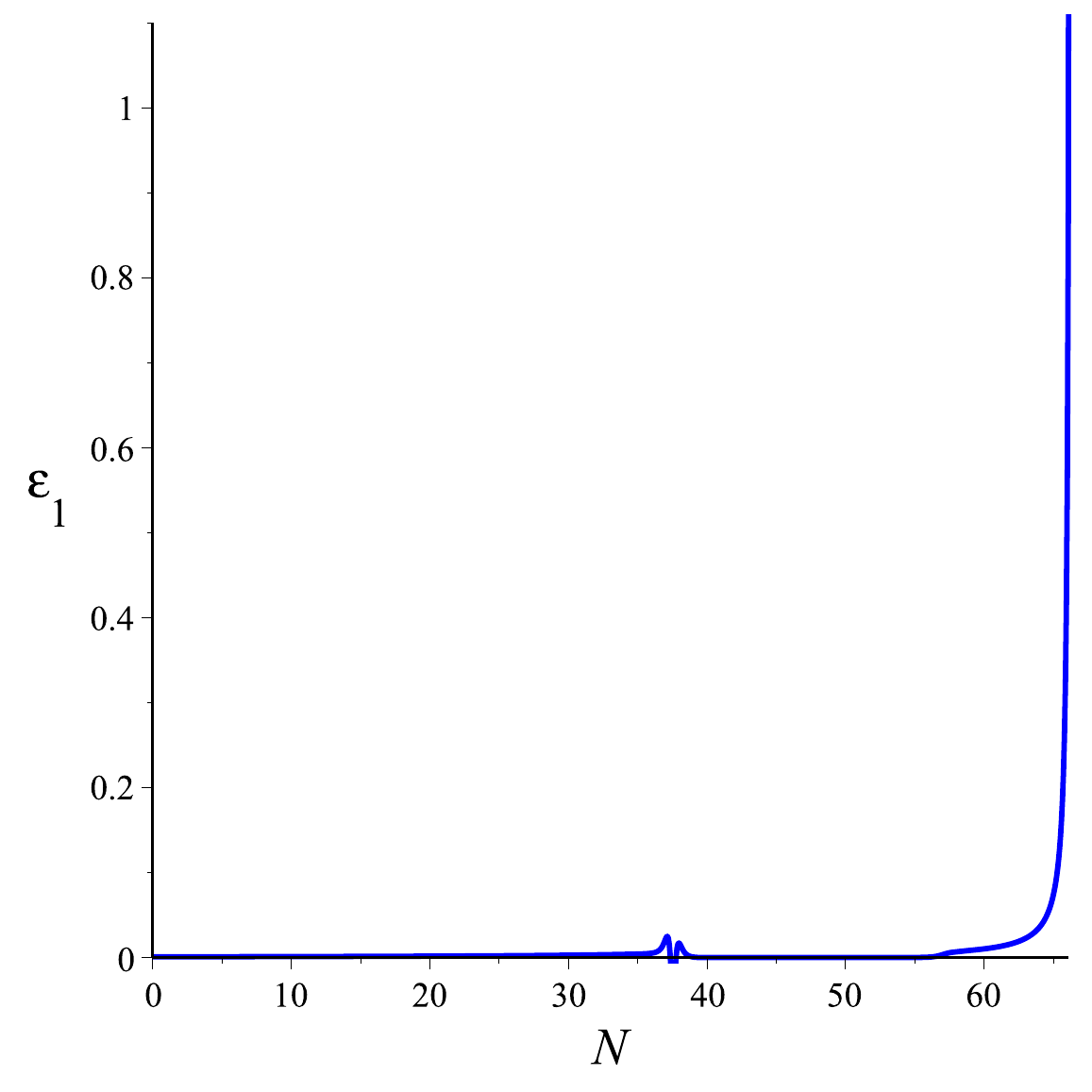}\quad
\includegraphics[scale=0.41]{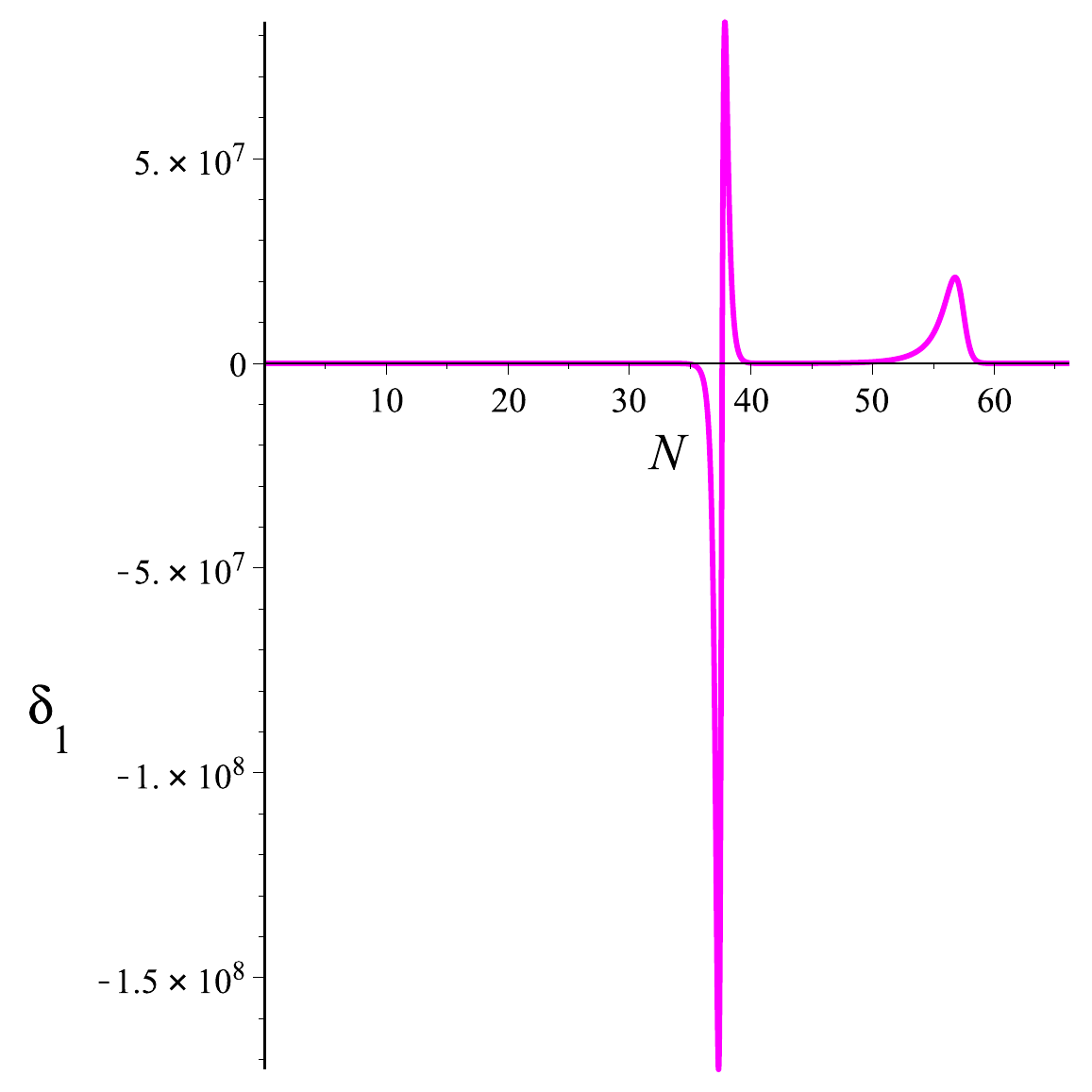}
\caption{The evolution of the slow-roll parameters $\varepsilon_1(N)$ and $\delta_1(N)$ during
inflation calculated by numerical integration of system (\ref{DynSYSN}). The parameter $\xi_0=-0.00721915$, values of other parameters given by (\ref{param}). \label{Fig2}}
\end{figure}

\begin{figure}
\includegraphics[scale=0.41]{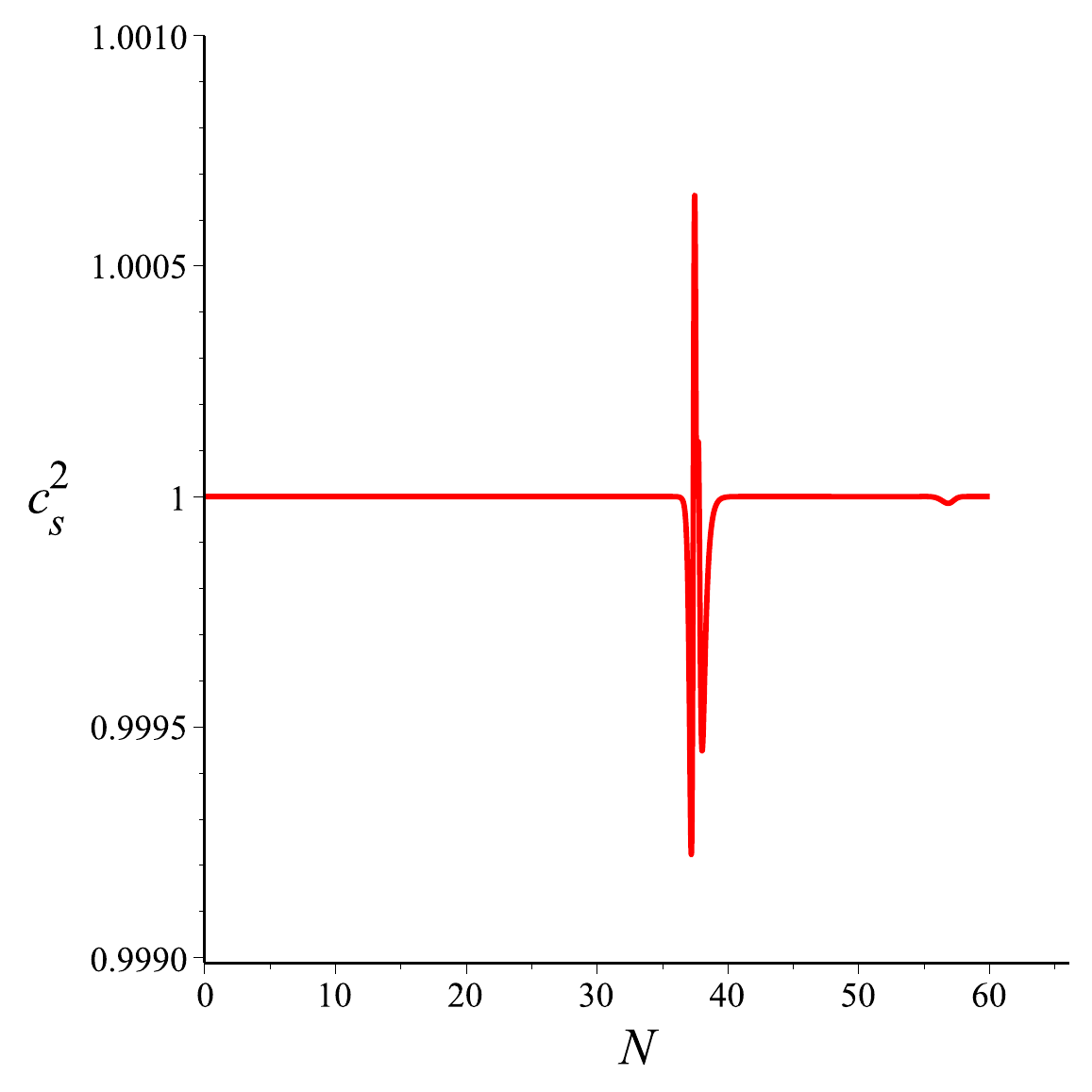}\quad
\includegraphics[scale=0.41]{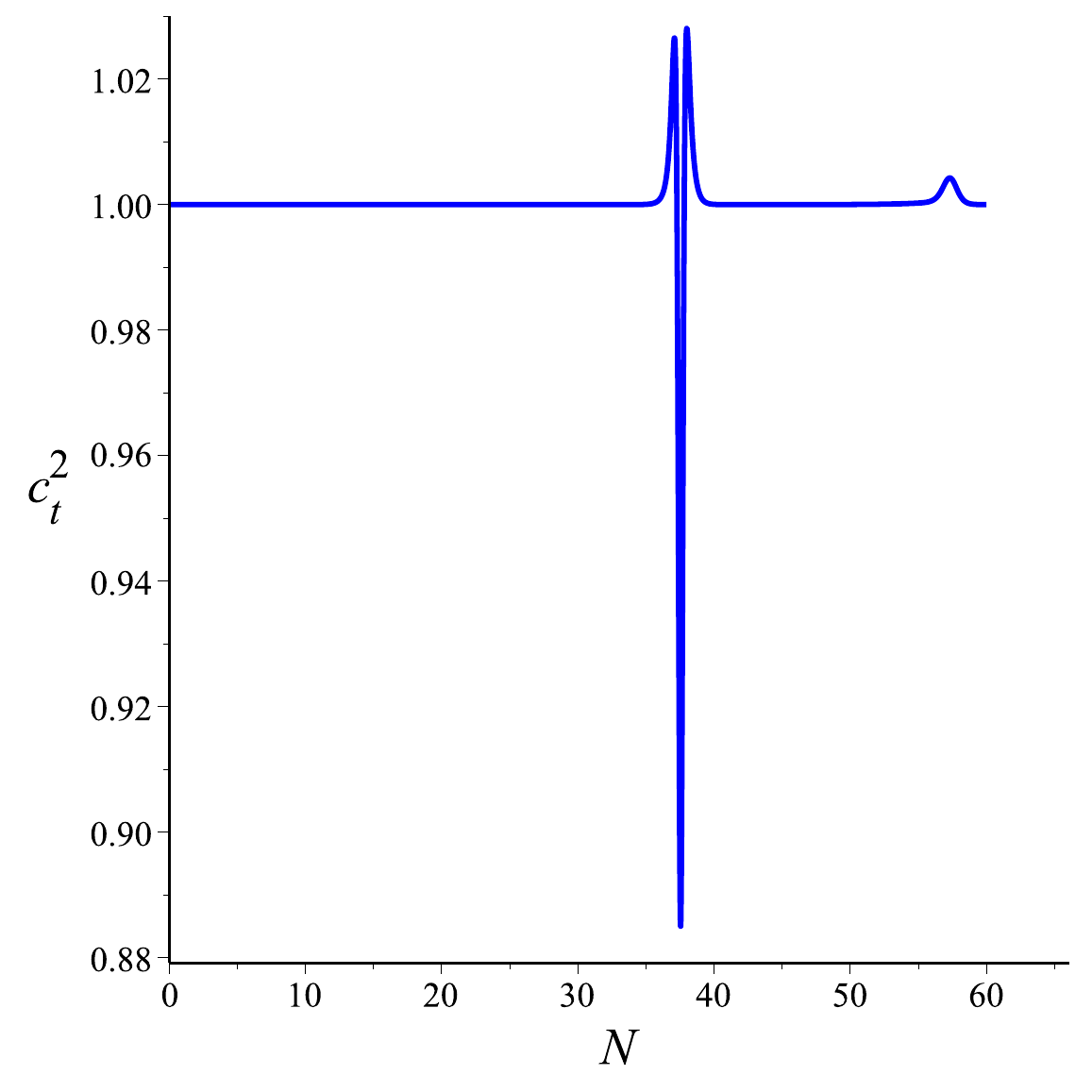}
\caption{The scalar perturbation (left) and tensor perturbation (right) propagation speeds during
inflation. The evolution is calculated by numerical integration of
system (\ref{DynSYSN}). The parameter $\xi_0=-0.00721915$, values of other parameters given by (\ref{param}). \label{Fig3}}
\end{figure}

The slow-roll parameter $\varepsilon_1$ is always less than $1$ during inflation, while the absolute value of $\delta_1$ is smaller than $1$ at the beginning and at the end of inflation only  (see Fig.~\ref{Fig2}). Therefore, the slow-roll conditions are violated during inflation. In Fig.~\ref{Fig3}, we demonstrate that the propagation squared speeds of scalar and tensor perturbations remain positive during inflation. So, the proposed model have no problem with stability.

\begin{figure}\label{AsStudy}
\includegraphics[scale=0.41]{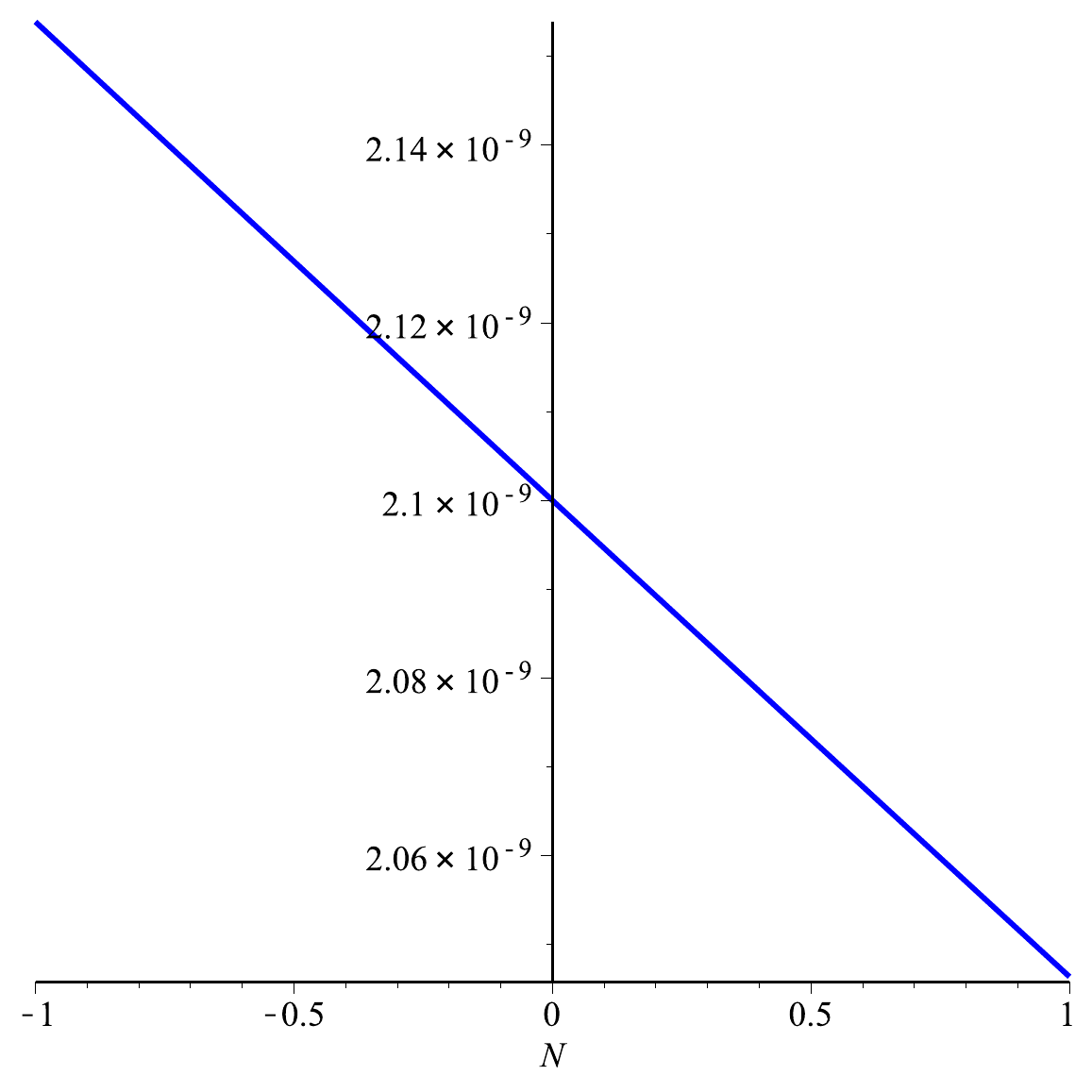}\quad
\includegraphics[scale=0.41]{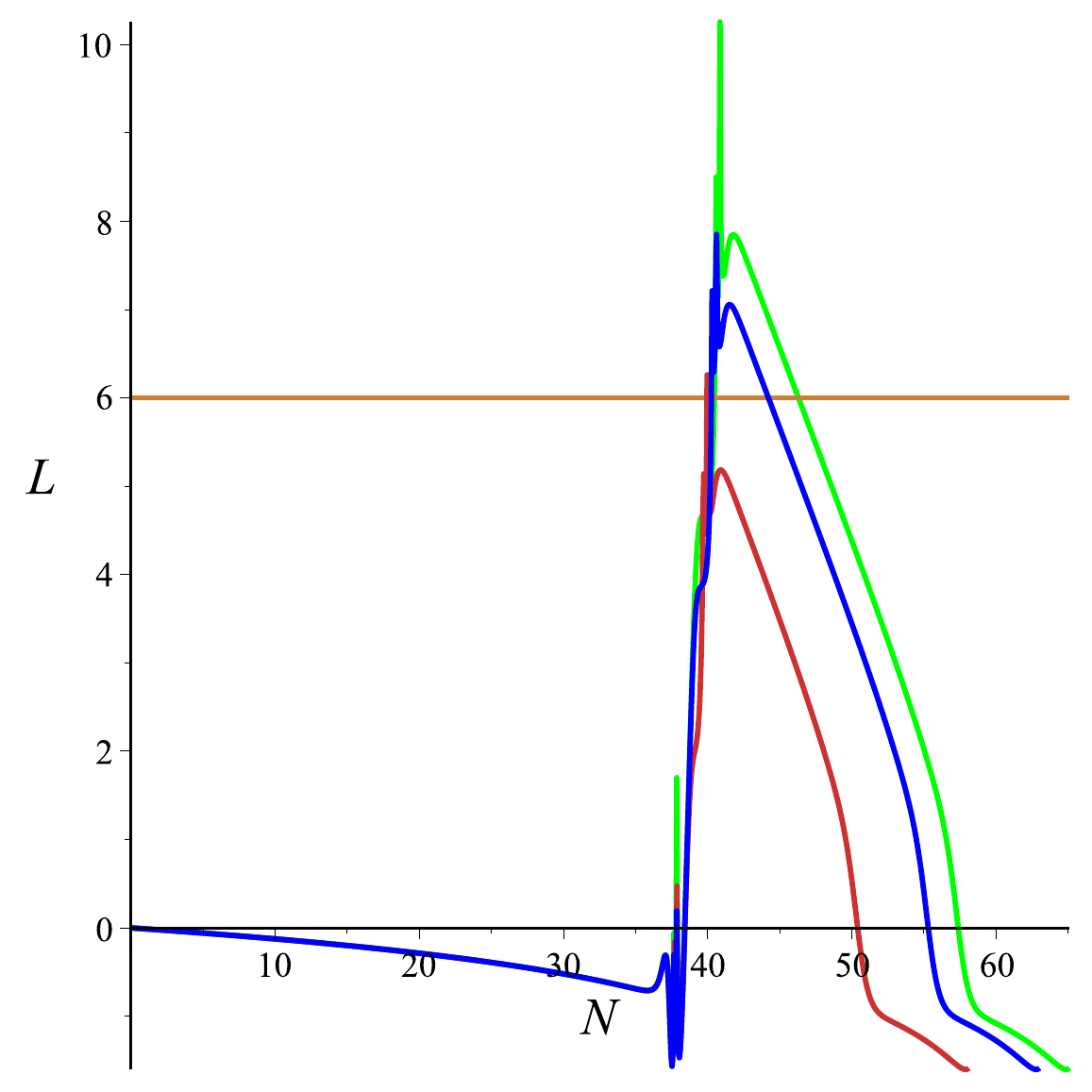}
\caption{The scalar perturbation amplitude ${\cal P}_{\zeta}$ at the beginning of inflation at $\xi_0=-0.00721915$ is presented in the left picture. 
In the right picture, the function $L=Log_{10}\left({\cal
P}_{\zeta}/{\cal P}_{\zeta,CMB}\right)$ is presented. The green curve corresponds to
$\xi_0=-0.00721915$, the blue curve corresponds to $\xi_0=-0.007219$,
the orange curve corresponds to $\xi_0=-0.0072165$. The evolution is
calculated by numerical integration of system (\ref{DynSYSN}). The
parameter values of other parameters given by (\ref{param}).
\label{Fig4}}
\end{figure}

In Fig.~\ref{Fig4}, we present the function ${\cal P}_{\zeta}$ at the beginning of inflation and the function $L$ from the beginning to end of inflation. The existence of some interval in $N$ where the function $L>6$ is necessary for PBH formation. At $\xi_0=-0.0072165$, this interval is less than $1$, whereas  at $\xi_0=-0.007219$ and $\xi_0=-0.00721915$ this interval is more than $5$.

\section{CONCLUSION}

In this paper, we propose a new single-field inflationary model that is in good agreement with the ACT observational data. In the proposed model, the Gauss-Bonnet term plays an important role, because the corresponding model with the same potential and without Gauss-Bonnet term not only does not provides conditions for PBH production, but also contradicts to the observation data.
Both scalar and tensor perturbation propagation squared speeds are positive during inflation. To analyze the ultra-slow roll regime we use the effective potential and show that the approximate equation (\ref{USRequ}) has the same form as in models without the Gauss-Bonnet term, but the Gauss-Bonnet term changes the condition for the existence of the ultra-slow-roll regime.

In this model, the Gauss-Bonnet term is essential for achieving the ultra-slow-roll phase of inflation, which provides the conditions for PBH production. Note that the proposed model requires fine-tuning of the parameter $\xi_0$ (see Table~\ref{Tab}). This common problem for single-field inflationary models with the ultra-slow-roll phase~\cite{Cole:2023wyx} has been explored recently for single-field models without the Gauss-Bonnet term and it was shown that such primordial black hole models are not technically unnatural~\cite{Iovino:2025tcv}. The proposed arguments looks correct and for more complicated single-field models including the Einstein-Gauss-Bonnet gravity models as well, but this question requires an additional consideration.

There is another approach to unify inflation and PBH formation~\cite{Garcia-Bellido:1996mdl,Pi:2017gih,Gundhi:2020kzm,Braglia:2020eai,Cheong:2022gfc,Geller:2022nkr,Braglia:2022phb,Pozdeeva:2024hfq,Wang:2024vfv,Pozdeeva:2025wsl,Pozdeeva:2026jxb}.  In two-field inflationary models, so-called two-stage inflation scenarios are realized. In the first stage, one scalar field evolves, while the other remains almost constant. This stage satisfies the slow-roll conditions. The second stage of inflation corresponds to the evolution of the second field, and the slow-rolling regime is violated when the first inflationary stage ends and the second begins~\cite{Ketov:2021fww}. These models do not have any fine-tuning problems.
Similar evolution appears in $F(R)$ models with a scalar field~\cite{Pi:2017gih,Gundhi:2020kzm,Pozdeeva:2025wsl}. It would be interesting to generalize the proposed model, by introducing some nonlinear functions of either the Ricci scalar or the Gauss-Bonnet term.

\section*{FUNDING}
This study was conducted within the scientific program of the National Center for Physics and Mathematics, section 5 'Particle Physics and Cosmology'. Stage 2026--2027.

%\section*{CONFLICT OF INTEREST}

% The authors of this work declare that they have no conflicts of interest.

%\section{Data Availability Statement}
% No new data were created or analyzed in this study.

\bibliographystyle{utphys}
\bibliography{INSPIRE-GB_Inflation}{}

\providecommand{\href}[2]{#2}\begingroup\raggedright\begin{thebibliography}{100}

\bibitem{Metsaev:1986yb}
R.~R. Metsaev and A.~A. Tseytlin, ``{Curvature Cubed Terms in String Theory
  Effective Actions},''
  \href{http://dx.doi.org/10.1016/0370-2693(87)91527-9}{{\em Phys. Lett. B}
  {\bfseries 185} (1987) 52--58}.

\bibitem{Berkin:1990nu}
A.~L. Berkin and K.-i. Maeda, ``{Effects of R**3 and R box R terms on R**2
  inflation},'' \href{http://dx.doi.org/10.1016/0370-2693(90)90657-R}{{\em
  Phys. Lett. B} {\bfseries 245} (1990) 348--354}.

\bibitem{Kawai:1999pw}
S.~Kawai and J.~Soda, ``{Evolution of fluctuations during graceful exit in
  string cosmology},''
  \href{http://dx.doi.org/10.1016/S0370-2693(99)00736-4}{{\em Phys. Lett. B}
  {\bfseries 460} (1999) 41--46},
  \href{http://arxiv.org/abs/gr-qc/9903017}{{\ttfamily arXiv:gr-qc/9903017}}.

\bibitem{Cartier:2001is}
C.~Cartier, J.-c. Hwang, and E.~J. Copeland, ``{Evolution of cosmological
  perturbations in nonsingular string cosmologies},''
  \href{http://dx.doi.org/10.1103/PhysRevD.64.103504}{{\em Phys. Rev. D}
  {\bfseries 64} (2001) 103504},
  \href{http://arxiv.org/abs/astro-ph/0106197}{{\ttfamily
  arXiv:astro-ph/0106197}}.

\bibitem{Hwang:2005hb}
J.-c. Hwang and H.~Noh, ``{Classical evolution and quantum generation in
  generalized gravity theories including string corrections and tachyon:
  Unified analyses},'' \href{http://dx.doi.org/10.1103/PhysRevD.71.063536}{{\em
  Phys. Rev. D} {\bfseries 71} (2005) 063536},
  \href{http://arxiv.org/abs/gr-qc/0412126}{{\ttfamily arXiv:gr-qc/0412126}}.

\bibitem{Nojiri:2005vv}
S.~Nojiri, S.~D. Odintsov, and M.~Sasaki, ``{Gauss-Bonnet dark energy},''
  \href{http://dx.doi.org/10.1103/PhysRevD.71.123509}{{\em Phys. Rev. D}
  {\bfseries 71} (2005) 123509},
  \href{http://arxiv.org/abs/hep-th/0504052}{{\ttfamily arXiv:hep-th/0504052}}.

\bibitem{Sami:2005zc}
M.~Sami, A.~Toporensky, P.~V. Tretjakov, and S.~Tsujikawa, ``{The Fate of
  (phantom) dark energy universe with string curvature corrections},''
  \href{http://dx.doi.org/10.1016/j.physletb.2005.06.017}{{\em Phys. Lett. B}
  {\bfseries 619} (2005) 193},
  \href{http://arxiv.org/abs/hep-th/0504154}{{\ttfamily arXiv:hep-th/0504154}}.

\bibitem{Tsujikawa:2006ph}
S.~Tsujikawa and M.~Sami, ``{String-inspired cosmology: Late time transition
  from scaling matter era to dark energy universe caused by a Gauss-Bonnet
  coupling},'' \href{http://dx.doi.org/10.1088/1475-7516/2007/01/006}{{\em
  JCAP} {\bfseries 01} (2007) 006},
  \href{http://arxiv.org/abs/hep-th/0608178}{{\ttfamily arXiv:hep-th/0608178}}.

\bibitem{Cognola:2006sp}
G.~Cognola, E.~Elizalde, S.~Nojiri, S.~Odintsov, and S.~Zerbini,
  ``{String-inspired Gauss-Bonnet gravity reconstructed from the universe
  expansion history and yielding the transition from matter dominance to dark
  energy},'' \href{http://dx.doi.org/10.1103/PhysRevD.75.086002}{{\em Phys.
  Rev. D} {\bfseries 75} (2007) 086002},
  \href{http://arxiv.org/abs/hep-th/0611198}{{\ttfamily arXiv:hep-th/0611198}}.

\bibitem{Guo:2006ct}
Z.-K. Guo, N.~Ohta, and S.~Tsujikawa, ``{Realizing scale-invariant density
  perturbations in low-energy effective string theory},''
  \href{http://dx.doi.org/10.1103/PhysRevD.75.023520}{{\em Phys. Rev. D}
  {\bfseries 75} (2007) 023520},
  \href{http://arxiv.org/abs/hep-th/0610336}{{\ttfamily arXiv:hep-th/0610336}}.

\bibitem{Iihoshi:2007vv}
M.~Iihoshi and S.~V. Ketov, ``{On the superstrings-induced four-dimensional
  gravity, and its applications to cosmology},''
  \href{http://dx.doi.org/10.1155/2008/521389}{{\em Adv. High Energy Phys.}
  {\bfseries 2008} (2008) 521389},
  \href{http://arxiv.org/abs/0707.3359}{{\ttfamily arXiv:0707.3359 [hep-th]}}.

\bibitem{RomeroCastellanos:2018inv}
A.~R. Romero~Castellanos, F.~Sobreira, I.~L. Shapiro, and A.~A. Starobinsky,
  ``{On higher derivative corrections to the $R+R^2$ inflationary model},''
  \href{http://dx.doi.org/10.1088/1475-7516/2018/12/007}{{\em JCAP} {\bfseries
  12} (2018) 007}, \href{http://arxiv.org/abs/1810.07787}{{\ttfamily
  arXiv:1810.07787 [gr-qc]}}.

\bibitem{Ketov:2022lhx}
S.~V. Ketov, ``{Starobinsky{\textendash}Bel{\textendash}Robinson Gravity},''
  \href{http://dx.doi.org/10.3390/universe8070351}{{\em Universe} {\bfseries 8}
  no.~7, (2022) 351}, \href{http://arxiv.org/abs/2205.13172}{{\ttfamily
  arXiv:2205.13172 [gr-qc]}}.

\bibitem{Ketov:2022zhp}
S.~V. Ketov, E.~O. Pozdeeva, and S.~Y. Vernov, ``{On the superstring-inspired
  quantum correction to the Starobinsky model of inflation},''
  \href{http://dx.doi.org/10.1088/1475-7516/2022/12/032}{{\em JCAP} {\bfseries
  12} (2022) 032}, \href{http://arxiv.org/abs/2211.01546}{{\ttfamily
  arXiv:2211.01546 [gr-qc]}}.

\bibitem{Gitsis:2023cyw}
A.~Gitsis, K.-R. Revis, S.~A. Venikoudis, and F.~P. Fronimos, ``{Swampland
  criteria for rescaled Einstein-Hilbert gravity with string corrections},''
  \href{http://arxiv.org/abs/2301.08126}{{\ttfamily arXiv:2301.08126 [gr-qc]}}.

\bibitem{Odintsov:2026cxz}
S.~D. Odintsov, V.~K. Oikonomou, P.~Tsyba, O.~Razina, and D.~Rakhatov,
  ``{String-inspired Gauss-Bonnet Gravity Inflation and ACT},''
  \href{http://dx.doi.org/10.1016/j.dark.2026.102348}{{\em Phys. Dark Univ.}
  {\bfseries 52} (2026) 102348},
  \href{http://arxiv.org/abs/2604.18861}{{\ttfamily arXiv:2604.18861 [gr-qc]}}.

\bibitem{Guo:2010jr}
Z.-K. Guo and D.~J. Schwarz, ``{Slow-roll inflation with a Gauss-Bonnet
  correction},'' \href{http://dx.doi.org/10.1103/PhysRevD.81.123520}{{\em Phys.
  Rev. D} {\bfseries 81} (2010) 123520},
  \href{http://arxiv.org/abs/1001.1897}{{\ttfamily arXiv:1001.1897 [hep-th]}}.

\bibitem{Jiang:2013gza}
P.-X. Jiang, J.-W. Hu, and Z.-K. Guo, ``{Inflation coupled to a Gauss-Bonnet
  term},'' \href{http://dx.doi.org/10.1103/PhysRevD.88.123508}{{\em Phys. Rev.
  D} {\bfseries 88} (2013) 123508},
  \href{http://arxiv.org/abs/1310.5579}{{\ttfamily arXiv:1310.5579 [hep-th]}}.

\bibitem{Koh:2014bka}
S.~Koh, B.-H. Lee, W.~Lee, and G.~Tumurtushaa, ``{Observational constraints on
  slow-roll inflation coupled to a Gauss-Bonnet term},''
  \href{http://dx.doi.org/10.1103/PhysRevD.90.063527}{{\em Phys. Rev. D}
  {\bfseries 90} no.~6, (2014) 063527},
  \href{http://arxiv.org/abs/1404.6096}{{\ttfamily arXiv:1404.6096 [gr-qc]}}.

\bibitem{Hikmawan:2015rze}
G.~Hikmawan, J.~Soda, A.~Suroso, and F.~P. Zen, ``{Comment on
  \textquotedblleft{}Gauss-Bonnet inflation\textquotedblright{}},''
  \href{http://dx.doi.org/10.1103/PhysRevD.93.068301}{{\em Phys. Rev. D}
  {\bfseries 93} no.~6, (2016) 068301},
  \href{http://arxiv.org/abs/1512.00222}{{\ttfamily arXiv:1512.00222
  [hep-th]}}.

\bibitem{vandeBruck:2015gjd}
C.~van~de Bruck and C.~Longden, ``{Higgs Inflation with a Gauss-Bonnet term in
  the Jordan Frame},'' \href{http://dx.doi.org/10.1103/PhysRevD.93.063519}{{\em
  Phys. Rev. D} {\bfseries 93} no.~6, (2016) 063519},
  \href{http://arxiv.org/abs/1512.04768}{{\ttfamily arXiv:1512.04768
  [hep-ph]}}.

\bibitem{vandeBruck:2017voa}
C.~van~de Bruck, K.~Dimopoulos, C.~Longden, and C.~Owen,
  ``{Gauss-Bonnet-coupled Quintessential Inflation},''
  \href{http://arxiv.org/abs/1707.06839}{{\ttfamily arXiv:1707.06839
  [astro-ph.CO]}}.

\bibitem{Koh:2016abf}
S.~Koh, B.-H. Lee, and G.~Tumurtushaa, ``{Reconstruction of the Scalar Field
  Potential in Inflationary Models with a Gauss-Bonnet term},''
  \href{http://dx.doi.org/10.1103/PhysRevD.95.123509}{{\em Phys. Rev. D}
  {\bfseries 95} no.~12, (2017) 123509},
  \href{http://arxiv.org/abs/1610.04360}{{\ttfamily arXiv:1610.04360 [gr-qc]}}.

\bibitem{Mathew:2016anx}
J.~Mathew and S.~Shankaranarayanan, ``{Low scale Higgs inflation with
  Gauss{\textendash}Bonnet coupling},''
  \href{http://dx.doi.org/10.1016/j.astropartphys.2016.07.004}{{\em Astropart.
  Phys.} {\bfseries 84} (2016) 1--7},
  \href{http://arxiv.org/abs/1602.00411}{{\ttfamily arXiv:1602.00411
  [astro-ph.CO]}}.

\bibitem{Chakraborty:2018scm}
S.~Chakraborty, T.~Paul, and S.~SenGupta, ``{Inflation driven by
  Einstein-Gauss-Bonnet gravity},''
  \href{http://dx.doi.org/10.1103/PhysRevD.98.083539}{{\em Phys. Rev. D}
  {\bfseries 98} no.~8, (2018) 083539},
  \href{http://arxiv.org/abs/1804.03004}{{\ttfamily arXiv:1804.03004 [gr-qc]}}.

\bibitem{Yi:2018gse}
Z.~Yi, Y.~Gong, and M.~Sabir, ``{Inflation with Gauss-Bonnet coupling},''
  \href{http://dx.doi.org/10.1103/PhysRevD.98.083521}{{\em Phys. Rev. D}
  {\bfseries 98} no.~8, (2018) 083521},
  \href{http://arxiv.org/abs/1804.09116}{{\ttfamily arXiv:1804.09116 [gr-qc]}}.

\bibitem{Odintsov:2018zhw}
S.~D. Odintsov and V.~K. Oikonomou, ``{Viable Inflation in Scalar-Gauss-Bonnet
  Gravity and Reconstruction from Observational Indices},''
  \href{http://dx.doi.org/10.1103/PhysRevD.98.044039}{{\em Phys. Rev. D}
  {\bfseries 98} no.~4, (2018) 044039},
  \href{http://arxiv.org/abs/1808.05045}{{\ttfamily arXiv:1808.05045 [gr-qc]}}.

\bibitem{Kleidis:2019ywv}
K.~Kleidis and V.~K. Oikonomou, ``{A Study of an Einstein Gauss-Bonnet
  Quintessential Inflationary Model},''
  \href{http://dx.doi.org/10.1016/j.nuclphysb.2019.114765}{{\em Nucl. Phys. B}
  {\bfseries 948} (2019) 114765},
  \href{http://arxiv.org/abs/1909.05318}{{\ttfamily arXiv:1909.05318 [gr-qc]}}.

\bibitem{Odintsov:2019clh}
S.~D. Odintsov and V.~K. Oikonomou, ``{Inflationary Phenomenology of Einstein
  Gauss-Bonnet Gravity Compatible with GW170817},''
  \href{http://dx.doi.org/10.1016/j.physletb.2019.134874}{{\em Phys. Lett. B}
  {\bfseries 797} (2019) 134874},
  \href{http://arxiv.org/abs/1908.07555}{{\ttfamily arXiv:1908.07555 [gr-qc]}}.

\bibitem{Odintsov:2020zkl}
S.~D. Odintsov and V.~K. Oikonomou, ``{Swampland implications of
  GW170817-compatible Einstein-Gauss-Bonnet gravity},''
  \href{http://dx.doi.org/10.1016/j.physletb.2020.135437}{{\em Phys. Lett. B}
  {\bfseries 805} (2020) 135437},
  \href{http://arxiv.org/abs/2004.00479}{{\ttfamily arXiv:2004.00479 [gr-qc]}}.

\bibitem{Odintsov:2020sqy}
S.~D. Odintsov, V.~K. Oikonomou, and F.~P. Fronimos, ``{Rectifying
  Einstein-Gauss-Bonnet Inflation in View of GW170817},''
  \href{http://dx.doi.org/10.1016/j.nuclphysb.2020.115135}{{\em Nucl. Phys. B}
  {\bfseries 958} (2020) 115135},
  \href{http://arxiv.org/abs/2003.13724}{{\ttfamily arXiv:2003.13724 [gr-qc]}}.

\bibitem{Rashidi:2020wwg}
N.~Rashidi and K.~Nozari, ``{Gauss-Bonnet Inflation after Planck2018},''
  \href{http://dx.doi.org/10.3847/1538-4357/ab6a10}{{\em Astrophys. J.}
  {\bfseries 890} (1, 2020) 58},
  \href{http://arxiv.org/abs/2001.07012}{{\ttfamily arXiv:2001.07012
  [astro-ph.CO]}}.

\bibitem{Pozdeeva:2020shl}
E.~O. Pozdeeva, ``{Generalization of cosmological attractor approach to
  Einstein{\textendash}Gauss{\textendash}Bonnet gravity},''
  \href{http://dx.doi.org/10.1140/epjc/s10052-020-8176-3}{{\em Eur. Phys. J. C}
  {\bfseries 80} no.~7, (2020) 612},
  \href{http://arxiv.org/abs/2005.10133}{{\ttfamily arXiv:2005.10133 [gr-qc]}}.

\bibitem{Pozdeeva:2020apf}
E.~O. Pozdeeva, M.~R. Gangopadhyay, M.~Sami, A.~V. Toporensky, and S.~Y.
  Vernov, ``{Inflation with a quartic potential in the framework of
  Einstein-Gauss-Bonnet gravity},''
  \href{http://dx.doi.org/10.1103/PhysRevD.102.043525}{{\em Phys. Rev. D}
  {\bfseries 102} no.~4, (2020) 043525},
  \href{http://arxiv.org/abs/2006.08027}{{\ttfamily arXiv:2006.08027 [gr-qc]}}.

\bibitem{Pozdeeva:2021iwc}
E.~O. Pozdeeva and S.~Y. Vernov, ``{Construction of inflationary scenarios with
  the Gauss{\textendash}Bonnet term and nonminimal coupling},''
  \href{http://dx.doi.org/10.1140/epjc/s10052-021-09435-8}{{\em Eur. Phys. J.
  C} {\bfseries 81} no.~7, (2021) 633},
  \href{http://arxiv.org/abs/2104.04995}{{\ttfamily arXiv:2104.04995 [gr-qc]}}.

\bibitem{Oikonomou:2021kql}
V.~K. Oikonomou, ``{A refined Einstein{\textendash}Gauss{\textendash}Bonnet
  inflationary theoretical framework},''
  \href{http://dx.doi.org/10.1088/1361-6382/ac2168}{{\em Class. Quant. Grav.}
  {\bfseries 38} no.~19, (2021) 195025},
  \href{http://arxiv.org/abs/2108.10460}{{\ttfamily arXiv:2108.10460 [gr-qc]}}.

\bibitem{Pozdeeva:2021nmz}
E.~O. Pozdeeva, ``{Deviation from Slow-Roll Regime in the EGB Inflationary
  Models with r {\ensuremath{\sim}} Ne{\ensuremath{-}}1},''
  \href{http://dx.doi.org/10.3390/universe7060181}{{\em Universe} {\bfseries 7}
  no.~6, (2021) 181}, \href{http://arxiv.org/abs/2105.02772}{{\ttfamily
  arXiv:2105.02772 [gr-qc]}}.

\bibitem{Kawai:2021edk}
S.~Kawai and J.~Kim, ``{Primordial black holes from Gauss-Bonnet-corrected
  single field inflation},''
  \href{http://dx.doi.org/10.1103/PhysRevD.104.083545}{{\em Phys. Rev. D}
  {\bfseries 104} no.~8, (2021) 083545},
  \href{http://arxiv.org/abs/2108.01340}{{\ttfamily arXiv:2108.01340
  [astro-ph.CO]}}.

\bibitem{Odintsov:2022zrj}
S.~D. Odintsov, D.~Saez-Chillon~Gomez, and G.~S. Sharov, ``{Testing viable
  extensions of Einstein\textendash{}Gauss\textendash{}Bonnet gravity},''
  \href{http://dx.doi.org/10.1016/j.dark.2022.101100}{{\em Phys. Dark Univ.}
  {\bfseries 37} (2022) 101100},
  \href{http://arxiv.org/abs/2207.08513}{{\ttfamily arXiv:2207.08513 [gr-qc]}}.

\bibitem{Oikonomou:2022ksx}
V.~K. Oikonomou, P.~D. Katzanis, and I.~C. Papadimitriou, ``{Bottom-up
  reconstruction of viable GW170817 compatible
  Einstein\textendash{}Gauss\textendash{}Bonnet theories},''
  \href{http://dx.doi.org/10.1088/1361-6382/ac5eba}{{\em Class. Quant. Grav.}
  {\bfseries 39} no.~9, (2022) 095008},
  \href{http://arxiv.org/abs/2203.09867}{{\ttfamily arXiv:2203.09867 [gr-qc]}}.

\bibitem{Kawaguchi:2022nku}
R.~Kawaguchi and S.~Tsujikawa, ``{Primordial black holes from Higgs inflation
  with a Gauss-Bonnet coupling},''
  \href{http://dx.doi.org/10.1103/PhysRevD.107.063508}{{\em Phys. Rev. D}
  {\bfseries 107} no.~6, (2023) 063508},
  \href{http://arxiv.org/abs/2211.13364}{{\ttfamily arXiv:2211.13364
  [astro-ph.CO]}}.

\bibitem{Gangopadhyay:2022vgh}
M.~R. Gangopadhyay, H.~A. Khan, and Yogesh, ``{A case study of small field
  inflationary dynamics in the Einstein\textendash{}Gauss\textendash{}Bonnet
  framework in the light of GW170817},''
  \href{http://dx.doi.org/10.1016/j.dark.2023.101177}{{\em Phys. Dark Univ.}
  {\bfseries 40} (2023) 101177},
  \href{http://arxiv.org/abs/2205.15261}{{\ttfamily arXiv:2205.15261
  [astro-ph.CO]}}.

\bibitem{Venikoudis:2022gfg}
S.~A. Venikoudis and F.~P. Fronimos, ``{Reheating era in Gauss-Bonnet theories
  of gravity compatible with the GW170817 event},''
  \href{http://dx.doi.org/10.1016/j.nuclphysb.2022.115945}{{\em Nucl. Phys. B}
  {\bfseries 984} (2022) 115945},
  \href{http://arxiv.org/abs/2208.08333}{{\ttfamily arXiv:2208.08333 [gr-qc]}}.

\bibitem{Odintsov:2023aaw}
S.~D. Odintsov, V.~K. Oikonomou, and F.~P. Fronimos, ``{Inflationary Dynamics
  and Swampland Criteria for Modified Gauss-Bonnet Gravity Compatible with
  GW170817},'' \href{http://dx.doi.org/10.1103/PhysRevD.107.084007}{{\em Phys.
  Rev. D} {\bfseries 107} (2023) 08},
  \href{http://arxiv.org/abs/2303.14594}{{\ttfamily arXiv:2303.14594 [gr-qc]}}.

\bibitem{Nojiri:2023mvi}
S.~Nojiri, S.~D. Odintsov, and D.~S\'aez-Chill\'on~G\'omez, ``{Unifying
  inflation with early and late dark energy in
  Einstein\textendash{}Gauss\textendash{}Bonnet gravity},''
  \href{http://dx.doi.org/10.1016/j.dark.2023.101238}{{\em Phys. Dark Univ.}
  {\bfseries 41} (2023) 101238},
  \href{http://arxiv.org/abs/2304.08255}{{\ttfamily arXiv:2304.08255 [gr-qc]}}.

\bibitem{Ashrafzadeh:2023ndt}
A.~Ashrafzadeh and K.~Karami, ``{Primordial Black Holes in Scalar Field
  Inflation Coupled to the Gauss{\textendash}Bonnet Term with Fractional
  Power-law Potentials},''
  \href{http://dx.doi.org/10.3847/1538-4357/ad293f}{{\em Astrophys. J.}
  {\bfseries 965} no.~1, (2024) 11},
  \href{http://arxiv.org/abs/2309.16356}{{\ttfamily arXiv:2309.16356
  [astro-ph.CO]}}.

\bibitem{Odintsov:2023lbb}
S.~D. Odintsov and T.~Paul, ``{From inflation to reheating and their dynamical
  stability analysis in Gauss{\textendash}Bonnet gravity},''
  \href{http://dx.doi.org/10.1016/j.dark.2023.101263}{{\em Phys. Dark Univ.}
  {\bfseries 42} (2023) 101263},
  \href{http://arxiv.org/abs/2305.19110}{{\ttfamily arXiv:2305.19110 [gr-qc]}}.

\bibitem{Mudrunka:2023wxy}
K.~Mudrunka and K.~Nakayama, ``{Probing Gauss-Bonnet-corrected inflation with
  gravitational waves},''
  \href{http://dx.doi.org/10.1088/1475-7516/2024/05/069}{{\em JCAP} {\bfseries
  05} (2024) 069}, \href{http://arxiv.org/abs/2312.15766}{{\ttfamily
  arXiv:2312.15766 [astro-ph.CO]}}.

\bibitem{Koh:2023zgn}
S.~Koh, S.~C. Park, and G.~Tumurtushaa, ``{Higgs inflation with a Gauss-Bonnet
  term},'' \href{http://dx.doi.org/10.1103/PhysRevD.110.023523}{{\em Phys. Rev.
  D} {\bfseries 110} no.~2, (2024) 023523},
  \href{http://arxiv.org/abs/2308.00897}{{\ttfamily arXiv:2308.00897 [gr-qc]}}.

\bibitem{Nojiri:2024cal}
S.~Nojiri and G.~G.~L. Nashed, ``{Gravitational collapse, primordial black
  hole, and gravitational wave in Einstein-Gauss-Bonnet theory with two scalar
  fields},'' \href{http://dx.doi.org/10.1103/PhysRevD.110.084073}{{\em Phys.
  Rev. D} {\bfseries 110} no.~8, (2024) 084073},
  \href{http://arxiv.org/abs/2405.00936}{{\ttfamily arXiv:2405.00936 [gr-qc]}}.

\bibitem{Pozdeeva:2024kzb}
E.~O. Pozdeeva, ``{Slow-roll approximations in
  Einstein{\textendash}Gauss{\textendash}Bonnet gravity formulated in terms of
  e-folding numbers},''
  \href{http://dx.doi.org/10.1140/epjc/s10052-025-13895-7}{{\em Eur. Phys. J.
  C} {\bfseries 85} no.~2, (2025) 217},
  \href{http://arxiv.org/abs/2411.16194}{{\ttfamily arXiv:2411.16194 [gr-qc]}}.

\bibitem{Pozdeeva:2024ihc}
E.~O. Pozdeeva, M.~A. Skugoreva, A.~V. Toporensky, and S.~Y. Vernov, ``{New
  slow-roll approximations for inflation in Einstein-Gauss-Bonnet gravity},''
  \href{http://dx.doi.org/10.1088/1475-7516/2024/09/050}{{\em JCAP} {\bfseries
  09} (2024) 050}, \href{http://arxiv.org/abs/2403.06147}{{\ttfamily
  arXiv:2403.06147 [gr-qc]}}.

\bibitem{Oikonomou:2024etl}
V.~K. Oikonomou, ``{Revisiting Einstein-Gauss-Bonnet theories after
  GW170817},'' \href{http://dx.doi.org/10.1016/j.physletb.2024.138890}{{\em
  Phys. Lett. B} {\bfseries 856} (2024) 138890},
  \href{http://arxiv.org/abs/2407.12155}{{\ttfamily arXiv:2407.12155 [gr-qc]}}.

\bibitem{Ashrafzadeh:2024oll}
A.~Ashrafzadeh, M.~Solbi, S.~Heydari, and K.~Karami, ``{Primordial black holes
  in SB SUSY Gauss-Bonnet inflation},''
  \href{http://dx.doi.org/10.1088/1475-7516/2025/01/025}{{\em JCAP} {\bfseries
  01} (2025) 025}, \href{http://arxiv.org/abs/2407.15445}{{\ttfamily
  arXiv:2407.15445 [hep-th]}}.

\bibitem{Solbi:2024zhl}
M.~Solbi and K.~Karami, ``{Primordial black holes in non-minimal
  Gauss{\textendash}Bonnet inflation in light of the PTA data},''
  \href{http://dx.doi.org/10.1140/epjc/s10052-024-13271-x}{{\em Eur. Phys. J.
  C} {\bfseries 84} no.~9, (2024) 918},
  \href{http://arxiv.org/abs/2403.00021}{{\ttfamily arXiv:2403.00021 [gr-qc]}}.

\bibitem{Odintsov:2025wai}
S.~D. Odintsov and V.~K. Oikonomou, ``{GW170817 Viable Einstein-Gauss-Bonnet
  inflation compatible with the atacama cosmology telescope data},''
  \href{http://dx.doi.org/10.1016/j.physletb.2025.139779}{{\em Phys. Lett. B}
  {\bfseries 868} (2025) 139779},
  \href{http://arxiv.org/abs/2506.08193}{{\ttfamily arXiv:2506.08193 [gr-qc]}}.

\bibitem{Mudrunka:2025xcg}
K.~Mudrunka and K.~Nakayama, ``{Inflation with Gauss-Bonnet correction: beyond
  slow-roll},'' \href{http://dx.doi.org/10.1088/1475-7516/2026/02/070}{{\em
  JCAP} {\bfseries 02} (2026) 070},
  \href{http://arxiv.org/abs/2504.01365}{{\ttfamily arXiv:2504.01365 [gr-qc]}}.

\bibitem{Ahghari:2025hfy}
Z.~Ahghari and M.~Farhoudi, ``{Inflation with Gauss{\textendash}Bonnet
  correction and Higgs potential},''
  \href{http://dx.doi.org/10.1140/epjc/s10052-026-15854-2}{{\em Eur. Phys. J.
  C} {\bfseries 86} no.~6, (2026) 601},
  \href{http://arxiv.org/abs/2512.12286}{{\ttfamily arXiv:2512.12286 [gr-qc]}}.

\bibitem{Yogesh:2025hll}
Yogesh and A.~Mohammadi, ``{Nonstandard Thermal History and Formation of
  Primordial Black Holes and SIGWs in
  Einstein{\textendash}Gauss{\textendash}Bonnet Gravity},''
  \href{http://dx.doi.org/10.3847/1538-4357/adcee5}{{\em Astrophys. J.}
  {\bfseries 986} no.~1, (2025) 35},
  \href{http://arxiv.org/abs/2501.01867}{{\ttfamily arXiv:2501.01867 [gr-qc]}}.

\bibitem{Yogesh:2025wak}
Yogesh, A.~Mohammadi, Q.~Wu, and T.~Zhu, ``{Starobinsky like inflation and EGB
  Gravity in the light of ACT},''
  \href{http://dx.doi.org/10.1088/1475-7516/2025/10/010}{{\em JCAP} {\bfseries
  10} (2025) 010}, \href{http://arxiv.org/abs/2505.05363}{{\ttfamily
  arXiv:2505.05363 [astro-ph.CO]}}.

\bibitem{Aldabergenov:2025oys}
Y.~Aldabergenov and D.~Berkimbayev, ``{Gauss{\textendash}Bonnet-Induced
  Symmetry Breaking/Restoration During Inflation},''
  \href{http://dx.doi.org/10.3390/universe11030098}{{\em Universe} {\bfseries
  11} no.~3, (2025) 98}, \href{http://arxiv.org/abs/2502.08986}{{\ttfamily
  arXiv:2502.08986 [hep-th]}}.

\bibitem{Rudenko:2026wkb}
B.~A. Rudenko, M.~A. Skugoreva, and A.~V. Toporensky, ``{Slow-roll
  approximations for Gauss-Bonnet inflation revisited},''
  \href{http://arxiv.org/abs/2601.23256}{{\ttfamily arXiv:2601.23256 [gr-qc]}}.

\bibitem{Yogesh:2026esn}
Yogesh, I.~A. Bhat, M.~R. Gangopadhyay, and M.~Sami, ``{Constraining
  quintessential inflation with ACT: A Gauss-Bonnet gateway},''
  \href{http://dx.doi.org/10.1088/1475-7516/2026/09/064}{{\em JCAP} {\bfseries
  09} (2026) 064}, \href{http://arxiv.org/abs/2604.14659}{{\ttfamily
  arXiv:2604.14659 [astro-ph.CO]}}.

\bibitem{Oikonomou:2026qkj}
V.~K. Oikonomou, ``{Reconstructing ACT-compatible and GW170817-compatible
  Einstein{\textendash}Gauss{\textendash}Bonnet Inflation from the
  Observational Indices},''
  \href{http://dx.doi.org/10.1016/j.aop.2026.170547}{{\em Annals Phys.}
  {\bfseries 492} (2026) 170547},
  \href{http://arxiv.org/abs/2605.27727}{{\ttfamily arXiv:2605.27727 [gr-qc]}}.

\bibitem{Manucharyan:2026twj}
G.~D. Manucharyan, I.~V. Fomin, S.~K. Mishra, and B.~Mishra,
  ``{Einstein--Gauss--Bonnet Inflationary Cosmology in Phase-$\Theta$
  Formalism},'' \href{http://arxiv.org/abs/2607.08807}{{\ttfamily
  arXiv:2607.08807 [gr-qc]}}.

\bibitem{Odintsov:2026veu}
S.~D. Odintsov and V.~K. Oikonomou, ``{Adiabatic Perturbations in
  GW170817-Compatible Einstein-Gauss-Bonnet Inflation},''
  \href{http://arxiv.org/abs/2608.09783}{{\ttfamily arXiv:2608.09783 [gr-qc]}}.

\bibitem{LIGOScientific:2017ync}
{\bfseries LIGO Scientific, Virgo, Fermi GBM, INTEGRAL, IceCube, AstroSat
  Cadmium Zinc Telluride Imager Team, IPN, Insight-Hxmt, ANTARES, Swift, AGILE
  Team, 1M2H Team, Dark Energy Camera GW-EM, DES, DLT40, GRAWITA, Fermi-LAT,
  ATCA, ASKAP, Las Cumbres Observatory Group, OzGrav, DWF (Deeper Wider Faster
  Program), AST3, CAASTRO, VINROUGE, MASTER, J-GEM, GROWTH, JAGWAR,
  CaltechNRAO, TTU-NRAO, NuSTAR, Pan-STARRS, MAXI Team, TZAC Consortium, KU,
  Nordic Optical Telescope, ePESSTO, GROND, Texas Tech University, SALT Group,
  TOROS, BOOTES, MWA, CALET, IKI-GW Follow-up, H.E.S.S., LOFAR, LWA, HAWC,
  Pierre Auger, ALMA, Euro VLBI Team, Pi of Sky, Chandra Team at McGill
  University, DFN, ATLAS Telescopes, High Time Resolution Universe Survey,
  RIMAS, RATIR, SKA South Africa/MeerKAT} Collaboration, B.~P. Abbott {\em
  et~al.}, ``{Multi-messenger Observations of a Binary Neutron Star Merger},''
  \href{http://dx.doi.org/10.3847/2041-8213/aa91c9}{{\em Astrophys. J. Lett.}
  {\bfseries 848} no.~2, (2017) L12},
  \href{http://arxiv.org/abs/1710.05833}{{\ttfamily arXiv:1710.05833
  [astro-ph.HE]}}.

\bibitem{Kobayashi:2011nu}
T.~Kobayashi, M.~Yamaguchi, and J.~Yokoyama, ``{Generalized G-inflation:
  Inflation with the most general second-order field equations},''
  \href{http://dx.doi.org/10.1143/PTP.126.511}{{\em Prog. Theor. Phys.}
  {\bfseries 126} (2011) 511--529},
  \href{http://arxiv.org/abs/1105.5723}{{\ttfamily arXiv:1105.5723 [hep-th]}}.

\bibitem{AtacamaCosmologyTelescope:2025blo}
{\bfseries Atacama Cosmology Telescope} Collaboration, T.~Louis {\em et~al.},
  ``{The Atacama Cosmology Telescope: DR6 power spectra, likelihoods and
  {\ensuremath{\Lambda}}CDM parameters},''
  \href{http://dx.doi.org/10.1088/1475-7516/2025/11/062}{{\em JCAP} {\bfseries
  11} (2025) 062}, \href{http://arxiv.org/abs/2503.14452}{{\ttfamily
  arXiv:2503.14452 [astro-ph.CO]}}.

\bibitem{AtacamaCosmologyTelescope:2025nti}
{\bfseries Atacama Cosmology Telescope} Collaboration, E.~Calabrese {\em
  et~al.}, ``{The Atacama Cosmology Telescope: DR6 constraints on extended
  cosmological models},''
  \href{http://dx.doi.org/10.1088/1475-7516/2025/11/063}{{\em JCAP} {\bfseries
  11} (2025) 063}, \href{http://arxiv.org/abs/2503.14454}{{\ttfamily
  arXiv:2503.14454 [astro-ph.CO]}}.

\bibitem{DESI:2024mwx}
{\bfseries DESI} Collaboration, A.~G. Adame {\em et~al.}, ``{DESI 2024 VI:
  cosmological constraints from the measurements of baryon acoustic
  oscillations},'' \href{http://dx.doi.org/10.1088/1475-7516/2025/02/021}{{\em
  JCAP} {\bfseries 02} (2025) 021},
  \href{http://arxiv.org/abs/2404.03002}{{\ttfamily arXiv:2404.03002
  [astro-ph.CO]}}.

\bibitem{DESI:2025zgx}
{\bfseries DESI} Collaboration, M.~Abdul~Karim {\em et~al.}, ``{DESI DR2
  results. II. Measurements of baryon acoustic oscillations and cosmological
  constraints},'' \href{http://dx.doi.org/10.1103/tr6y-kpc6}{{\em Phys. Rev. D}
  {\bfseries 112} no.~8, (2025) 083515},
  \href{http://arxiv.org/abs/2503.14738}{{\ttfamily arXiv:2503.14738
  [astro-ph.CO]}}.

\bibitem{Planck:2018jri}
{\bfseries Planck} Collaboration, Y.~Akrami {\em et~al.}, ``{Planck 2018
  results. X. Constraints on inflation},''
  \href{http://dx.doi.org/10.1051/0004-6361/201833887}{{\em Astron. Astrophys.}
  {\bfseries 641} (2020) A10},
  \href{http://arxiv.org/abs/1807.06211}{{\ttfamily arXiv:1807.06211
  [astro-ph.CO]}}.

\bibitem{BICEP:2021xfz}
{\bfseries BICEP, Keck} Collaboration, P.~A.~R. Ade {\em et~al.}, ``{Improved
  Constraints on Primordial Gravitational Waves using Planck, WMAP, and
  BICEP/Keck Observations through the 2018 Observing Season},''
  \href{http://dx.doi.org/10.1103/PhysRevLett.127.151301}{{\em Phys. Rev.
  Lett.} {\bfseries 127} no.~15, (2021) 151301},
  \href{http://arxiv.org/abs/2110.00483}{{\ttfamily arXiv:2110.00483
  [astro-ph.CO]}}.

\bibitem{Galloni:2022mok}
G.~Galloni, N.~Bartolo, S.~Matarrese, M.~Migliaccio, A.~Ricciardone, and
  N.~Vittorio, ``{Updated constraints on amplitude and tilt of the tensor
  primordial spectrum},''
  \href{http://dx.doi.org/10.1088/1475-7516/2023/04/062}{{\em JCAP} {\bfseries
  04} (2023) 062}, \href{http://arxiv.org/abs/2208.00188}{{\ttfamily
  arXiv:2208.00188 [astro-ph.CO]}}.

\bibitem{Ivanov:1994pa}
P.~Ivanov, P.~Naselsky, and I.~Novikov, ``{Inflation and primordial black holes
  as dark matter},'' \href{http://dx.doi.org/10.1103/PhysRevD.50.7173}{{\em
  Phys. Rev. D} {\bfseries 50} (1994) 7173--7178}.

\bibitem{Khlopov:2008qy}
M.~Y. Khlopov, ``{Primordial Black Holes},''
  \href{http://dx.doi.org/10.1088/1674-4527/10/6/001}{{\em Res. Astron.
  Astrophys.} {\bfseries 10} (2010) 495--528},
  \href{http://arxiv.org/abs/0801.0116}{{\ttfamily arXiv:0801.0116
  [astro-ph]}}.

\bibitem{Kamenshchik:2018sig}
A.~Y. Kamenshchik, A.~Tronconi, T.~Vardanyan, and G.~Venturi, ``{Non-Canonical
  Inflation and Primordial Black Holes Production},''
  \href{http://dx.doi.org/10.1016/j.physletb.2019.02.036}{{\em Phys. Lett. B}
  {\bfseries 791} (2019) 201--205},
  \href{http://arxiv.org/abs/1812.02547}{{\ttfamily arXiv:1812.02547 [gr-qc]}}.

\bibitem{Carr:2020xqk}
B.~Carr and F.~Kuhnel, ``{Primordial Black Holes as Dark Matter: Recent
  Developments},''
  \href{http://dx.doi.org/10.1146/annurev-nucl-050520-125911}{{\em Ann. Rev.
  Nucl. Part. Sci.} {\bfseries 70} (2020) 355--394},
  \href{http://arxiv.org/abs/2006.02838}{{\ttfamily arXiv:2006.02838
  [astro-ph.CO]}}.

\bibitem{Green:2020jor}
A.~M. Green and B.~J. Kavanagh, ``{Primordial Black Holes as a dark matter
  candidate},'' \href{http://dx.doi.org/10.1088/1361-6471/abc534}{{\em J. Phys.
  G} {\bfseries 48} no.~4, (2021) 043001},
  \href{http://arxiv.org/abs/2007.10722}{{\ttfamily arXiv:2007.10722
  [astro-ph.CO]}}.

\bibitem{Carr:2020gox}
B.~Carr, K.~Kohri, Y.~Sendouda, and J.~Yokoyama, ``{Constraints on primordial
  black holes},'' \href{http://dx.doi.org/10.1088/1361-6633/ac1e31}{{\em Rept.
  Prog. Phys.} {\bfseries 84} no.~11, (2021) 116902},
  \href{http://arxiv.org/abs/2002.12778}{{\ttfamily arXiv:2002.12778
  [astro-ph.CO]}}.

\bibitem{Ozsoy:2023ryl}
O.~{\"O}zsoy and G.~Tasinato, ``{Inflation and Primordial Black Holes},''
  \href{http://dx.doi.org/10.3390/universe9050203}{{\em Universe} {\bfseries 9}
  no.~5, (2023) 203}, \href{http://arxiv.org/abs/2301.03600}{{\ttfamily
  arXiv:2301.03600 [astro-ph.CO]}}.

\bibitem{Dolgov:2017aec}
A.~D. Dolgov, ``{Massive and supermassive black holes in the contemporary and
  early Universe and problems in cosmology and astrophysics},''
  \href{http://dx.doi.org/10.3367/UFNe.2017.06.038153}{{\em Usp. Fiz. Nauk}
  {\bfseries 188} no.~2, (2018) 121--142},
  \href{http://arxiv.org/abs/1705.06859}{{\ttfamily arXiv:1705.06859
  [astro-ph.CO]}}.

\bibitem{Dolgov:2020sov}
A.~Dolgov and K.~Postnov, ``{Why the mean mass of primordial black hole
  distribution is close to 10$M_{\odot}$},''
  \href{http://dx.doi.org/10.1088/1475-7516/2020/07/063}{{\em JCAP} {\bfseries
  07} (2020) 063}, \href{http://arxiv.org/abs/2004.11669}{{\ttfamily
  arXiv:2004.11669 [astro-ph.CO]}}.

\bibitem{Sasaki:2018dmp}
M.~Sasaki, T.~Suyama, T.~Tanaka, and S.~Yokoyama, ``{Primordial black
  holes{\textemdash}perspectives in gravitational wave astronomy},''
  \href{http://dx.doi.org/10.1088/1361-6382/aaa7b4}{{\em Class. Quant. Grav.}
  {\bfseries 35} no.~6, (2018) 063001},
  \href{http://arxiv.org/abs/1801.05235}{{\ttfamily arXiv:1801.05235
  [astro-ph.CO]}}.

\bibitem{DeLuca:2025fln}
V.~De~Luca, G.~Franciolini, and A.~Riotto, ``{GW231123: A Possible Primordial
  Black Hole Origin},'' \href{http://dx.doi.org/10.1103/y4x3-v6j2}{{\em Phys.
  Rev. Lett.} {\bfseries 136} no.~20, (2026) 201401},
  \href{http://arxiv.org/abs/2508.09965}{{\ttfamily arXiv:2508.09965
  [astro-ph.CO]}}.

\bibitem{Dolgov:1992pu}
A.~Dolgov and J.~Silk, ``{Baryon isocurvature fluctuations at small scales and
  baryonic dark matter},''
  \href{http://dx.doi.org/10.1103/PhysRevD.47.4244}{{\em Phys. Rev. D}
  {\bfseries 47} (1993) 4244--4255}.

\bibitem{Garcia-Bellido:1996mdl}
J.~Garcia-Bellido, A.~D. Linde, and D.~Wands, ``{Density perturbations and
  black hole formation in hybrid inflation},''
  \href{http://dx.doi.org/10.1103/PhysRevD.54.6040}{{\em Phys. Rev. D}
  {\bfseries 54} (1996) 6040--6058},
  \href{http://arxiv.org/abs/astro-ph/9605094}{{\ttfamily
  arXiv:astro-ph/9605094}}.

\bibitem{Ketov:2021fww}
S.~V. Ketov, ``{Multi-Field versus Single-Field in the Supergravity Models of
  Inflation and Primordial Black Holes},''
  \href{http://dx.doi.org/10.3390/universe7050115}{{\em Universe} {\bfseries 7}
  no.~5, (2021) 115}.

\bibitem{Garcia-Bellido:2017mdw}
J.~Garcia-Bellido and E.~Ruiz~Morales, ``{Primordial black holes from single
  field models of inflation},''
  \href{http://dx.doi.org/10.1016/j.dark.2017.09.007}{{\em Phys. Dark Univ.}
  {\bfseries 18} (2017) 47--54},
  \href{http://arxiv.org/abs/1702.03901}{{\ttfamily arXiv:1702.03901
  [astro-ph.CO]}}.

\bibitem{Germani:2017bcs}
C.~Germani and T.~Prokopec, ``{On primordial black holes from an inflection
  point},'' \href{http://dx.doi.org/10.1016/j.dark.2017.09.001}{{\em Phys. Dark
  Univ.} {\bfseries 18} (2017) 6--10},
  \href{http://arxiv.org/abs/1706.04226}{{\ttfamily arXiv:1706.04226
  [astro-ph.CO]}}.

\bibitem{Kannike:2017bxn}
K.~Kannike, L.~Marzola, M.~Raidal, and H.~Veerm{\"a}e, ``{Single Field Double
  Inflation and Primordial Black Holes},''
  \href{http://dx.doi.org/10.1088/1475-7516/2017/09/020}{{\em JCAP} {\bfseries
  09} (2017) 020}, \href{http://arxiv.org/abs/1705.06225}{{\ttfamily
  arXiv:1705.06225 [astro-ph.CO]}}.

\bibitem{Ezquiaga:2017fvi}
J.~M. Ezquiaga, J.~Garcia-Bellido, and E.~Ruiz~Morales, ``{Primordial Black
  Hole production in Critical Higgs Inflation},''
  \href{http://dx.doi.org/10.1016/j.physletb.2017.11.039}{{\em Phys. Lett. B}
  {\bfseries 776} (2018) 345--349},
  \href{http://arxiv.org/abs/1705.04861}{{\ttfamily arXiv:1705.04861
  [astro-ph.CO]}}.

\bibitem{Kristiano:2022maq}
J.~Kristiano and J.~Yokoyama, ``{Constraining Primordial Black Hole Formation
  from Single-Field Inflation},''
  \href{http://dx.doi.org/10.1103/PhysRevLett.132.221003}{{\em Phys. Rev.
  Lett.} {\bfseries 132} no.~22, (2024) 221003},
  \href{http://arxiv.org/abs/2211.03395}{{\ttfamily arXiv:2211.03395
  [hep-th]}}.

\bibitem{Chataignier:2023ago}
L.~Chataignier, A.~Y. Kamenshchik, A.~Tronconi, and G.~Venturi,
  ``{Reconstruction methods and the amplification of the inflationary
  spectrum},'' \href{http://dx.doi.org/10.1103/PhysRevD.107.083506}{{\em Phys.
  Rev. D} {\bfseries 107} no.~8, (2023) 083506},
  \href{http://arxiv.org/abs/2301.04477}{{\ttfamily arXiv:2301.04477 [gr-qc]}}.

\bibitem{Kamenshchik:2024kay}
A.~Y. Kamenshchik, E.~O. Pozdeeva, A.~Tribolet, A.~Tronconi, G.~Venturi, and
  S.~Y. Vernov, ``{Superpotential method and the amplification of inflationary
  perturbations},'' \href{http://dx.doi.org/10.1103/PhysRevD.110.104011}{{\em
  Phys. Rev. D} {\bfseries 110} no.~10, (2024) 104011},
  \href{http://arxiv.org/abs/2406.19762}{{\ttfamily arXiv:2406.19762 [gr-qc]}}.

\bibitem{Saburov:2023buy}
S.~Saburov and S.~V. Ketov, ``{Improved Model of Primordial Black Hole
  Formation after Starobinsky Inflation},''
  \href{http://dx.doi.org/10.3390/universe9070323}{{\em Universe} {\bfseries 9}
  no.~7, (2023) 323}, \href{http://arxiv.org/abs/2306.06597}{{\ttfamily
  arXiv:2306.06597 [gr-qc]}}.

\bibitem{Saburov:2024und}
S.~Saburov and S.~V. Ketov, ``{Quantum Loop Corrections in the Modified Gravity
  Model of Starobinsky Inflation with Primordial Black Hole Production},''
  \href{http://dx.doi.org/10.3390/universe10090354}{{\em Universe} {\bfseries
  10} no.~9, (2024) 354}, \href{http://arxiv.org/abs/2402.02934}{{\ttfamily
  arXiv:2402.02934 [gr-qc]}}.

\bibitem{Ketov:2025cqg}
S.~V. Ketov, E.~O. Pozdeeva, and S.~Y. Vernov, ``{Inflation in F(R) gravity
  models revisited after ACT},''
  \href{http://dx.doi.org/10.1088/1475-7516/2025/12/040}{{\em JCAP} {\bfseries
  12} (2025) 040}, \href{http://arxiv.org/abs/2508.08927}{{\ttfamily
  arXiv:2508.08927 [gr-qc]}}.

\bibitem{Iovino:2025tcv}
A.~J. Iovino and A.~Riotto, ``{Are primordial black holes truly fine-tuned?},''
  \href{http://dx.doi.org/10.1088/1475-7516/2026/07/047}{{\em JCAP} {\bfseries
  07} (2026) 047}, \href{http://arxiv.org/abs/2512.19668}{{\ttfamily
  arXiv:2512.19668 [astro-ph.CO]}}.

\bibitem{Totolou:2025dsb}
D.~Totolou, T.~Papanikolaou, and E.~N. Saridakis, ``{Primordial black hole dark
  matter from ultraslow-roll inflation in Horndeski gravity},''
  \href{http://dx.doi.org/10.1103/sfqc-q4tb}{{\em Phys. Rev. D} {\bfseries 114}
  no.~4, (2026) 044051}, \href{http://arxiv.org/abs/2512.25044}{{\ttfamily
  arXiv:2512.25044 [gr-qc]}}.

\bibitem{Pozdeeva:2019agu}
E.~O. Pozdeeva, M.~Sami, A.~V. Toporensky, and S.~Y. Vernov, ``{Stability
  analysis of de Sitter solutions in models with the Gauss-Bonnet term},''
  \href{http://dx.doi.org/10.1103/PhysRevD.100.083527}{{\em Phys. Rev. D}
  {\bfseries 100} no.~8, (2019) 083527},
  \href{http://arxiv.org/abs/1905.05085}{{\ttfamily arXiv:1905.05085 [gr-qc]}}.

\bibitem{Vernov:2021hxo}
S.~Vernov and E.~Pozdeeva, ``{De Sitter Solutions in
  Einstein{\textendash}Gauss{\textendash}Bonnet Gravity},''
  \href{http://dx.doi.org/10.3390/universe7050149}{{\em Universe} {\bfseries 7}
  no.~5, (2021) 149}, \href{http://arxiv.org/abs/2104.11111}{{\ttfamily
  arXiv:2104.11111 [gr-qc]}}.

\bibitem{Granda:2019wpe}
L.~N. Granda and D.~F. Jimenez, ``{Slow-Roll Inflation in Scalar-Tensor
  Models},'' \href{http://dx.doi.org/10.1088/1475-7516/2019/09/007}{{\em JCAP}
  {\bfseries 09} (2019) 007}, \href{http://arxiv.org/abs/1905.08349}{{\ttfamily
  arXiv:1905.08349 [gr-qc]}}.

\bibitem{Yogesh:2024mpa}
Yogesh, I.~A. Bhat, and M.~R. Gangopadhyay, ``{Inflationary dynamics in
  Einstein{\textendash}Gauss{\textendash}Bonnet gravity using new slow-roll
  approximations considering generalized reheating},''
  \href{http://dx.doi.org/10.1016/j.dark.2025.102021}{{\em Phys. Dark Univ.}
  {\bfseries 49} (2025) 102021},
  \href{http://arxiv.org/abs/2408.01670}{{\ttfamily arXiv:2408.01670
  [astro-ph.CO]}}.

\bibitem{Pi:2017gih}
S.~Pi, Y.-l. Zhang, Q.-G. Huang, and M.~Sasaki, ``{Scalaron from $R^2$-gravity
  as a heavy field},''
  \href{http://dx.doi.org/10.1088/1475-7516/2018/05/042}{{\em JCAP} {\bfseries
  05} (2018) 042}, \href{http://arxiv.org/abs/1712.09896}{{\ttfamily
  arXiv:1712.09896 [astro-ph.CO]}}.

\bibitem{Frolovsky:2022qpg}
D.~Frolovsky, S.~V. Ketov, and S.~Saburov, ``{E-models of inflation and
  primordial black holes},''
  \href{http://dx.doi.org/10.3389/fphy.2022.1005333}{{\em Front. in Phys.}
  {\bfseries 10} (2022) 1005333},
  \href{http://arxiv.org/abs/2207.11878}{{\ttfamily arXiv:2207.11878
  [astro-ph.CO]}}.

\bibitem{Pozdeeva:2024hfq}
E.~O. Pozdeeva and S.~Y. Vernov, ``{Primordial Black Holes in Induced Gravity
  Inflationary Models with Two Scalar Fields},''
  \href{http://dx.doi.org/10.1134/S1063779624701934}{{\em Phys. Part. Nucl.}
  {\bfseries 56} no.~2, (2025) 542--547},
  \href{http://arxiv.org/abs/2407.00999}{{\ttfamily arXiv:2407.00999 [gr-qc]}}.

\bibitem{Cole:2023wyx}
P.~S. Cole, A.~D. Gow, C.~T. Byrnes, and S.~P. Patil, ``{Primordial black holes
  from single-field inflation: a fine-tuning audit},''
  \href{http://dx.doi.org/10.1088/1475-7516/2023/08/031}{{\em JCAP} {\bfseries
  08} (2023) 031}, \href{http://arxiv.org/abs/2304.01997}{{\ttfamily
  arXiv:2304.01997 [astro-ph.CO]}}.

\bibitem{Gundhi:2020kzm}
A.~Gundhi, S.~V. Ketov, and C.~F. Steinwachs, ``{Primordial black hole dark
  matter in dilaton-extended two-field Starobinsky inflation},''
  \href{http://dx.doi.org/10.1103/PhysRevD.103.083518}{{\em Phys. Rev. D}
  {\bfseries 103} no.~8, (2021) 083518},
  \href{http://arxiv.org/abs/2011.05999}{{\ttfamily arXiv:2011.05999
  [hep-th]}}.

\bibitem{Braglia:2020eai}
M.~Braglia, D.~K. Hazra, F.~Finelli, G.~F. Smoot, L.~Sriramkumar, and A.~A.
  Starobinsky, ``{Generating PBHs and small-scale GWs in two-field models of
  inflation},'' \href{http://dx.doi.org/10.1088/1475-7516/2020/08/001}{{\em
  JCAP} {\bfseries 08} (2020) 001},
  \href{http://arxiv.org/abs/2005.02895}{{\ttfamily arXiv:2005.02895
  [astro-ph.CO]}}.

\bibitem{Cheong:2022gfc}
D.~Y. Cheong, K.~Kohri, and S.~C. Park, ``{The inflaton that could: primordial
  black holes and second order gravitational waves from tachyonic instability
  induced in Higgs-R $^{2}$ inflation},''
  \href{http://dx.doi.org/10.1088/1475-7516/2022/10/015}{{\em JCAP} {\bfseries
  10} (2022) 015}, \href{http://arxiv.org/abs/2205.14813}{{\ttfamily
  arXiv:2205.14813 [hep-ph]}}.

\bibitem{Geller:2022nkr}
S.~R. Geller, W.~Qin, E.~McDonough, and D.~I. Kaiser, ``{Primordial black holes
  from multifield inflation with nonminimal couplings},''
  \href{http://dx.doi.org/10.1103/PhysRevD.106.063535}{{\em Phys. Rev. D}
  {\bfseries 106} no.~6, (2022) 063535},
  \href{http://arxiv.org/abs/2205.04471}{{\ttfamily arXiv:2205.04471
  [hep-th]}}.

\bibitem{Braglia:2022phb}
M.~Braglia, A.~Linde, R.~Kallosh, and F.~Finelli, ``{Hybrid
  {\ensuremath{\alpha}}-attractors, primordial black holes and gravitational
  wave backgrounds},''
  \href{http://dx.doi.org/10.1088/1475-7516/2023/04/033}{{\em JCAP} {\bfseries
  04} (2023) 033}, \href{http://arxiv.org/abs/2211.14262}{{\ttfamily
  arXiv:2211.14262 [astro-ph.CO]}}.

\bibitem{Wang:2024vfv}
X.~Wang, Y.-l. Zhang, and M.~Sasaki, ``{Enhanced curvature perturbation and
  primordial black hole formation in two-stage inflation with a break},''
  \href{http://dx.doi.org/10.1088/1475-7516/2024/07/076}{{\em JCAP} {\bfseries
  07} (2024) 076}, \href{http://arxiv.org/abs/2404.02492}{{\ttfamily
  arXiv:2404.02492 [astro-ph.CO]}}.

\bibitem{Pozdeeva:2025wsl}
E.~O. Pozdeeva and S.~Y. Vernov, ``{Primordial Black Holes Formation in
  Inflationary $\boldsymbol{F(R)}$ Models with Scalar Fields},''
  \href{http://dx.doi.org/10.3103/S0027134925702753}{{\em Moscow Univ. Phys.
  Bull.} {\bfseries 80} no.~Suppl 2, (2025) S903--S912},
  \href{http://arxiv.org/abs/2509.21220}{{\ttfamily arXiv:2509.21220 [gr-qc]}}.

\bibitem{Pozdeeva:2026jxb}
E.~Pozdeeva, T.~Pozdeeva, and S.~Vernov, ``{Primordial Black Hole Production in
  Chiral Cosmological Inflationary Models},''
  \href{http://dx.doi.org/10.3390/psf2026014003}{{\em Phys. Sci. Forum}
  {\bfseries 14} no.~1, (2026) 3}.

\end{thebibliography}\endgroup

\end{document}